\documentclass[sigconf]{acmart}

\usepackage{tikz}

\usepackage{filecontents}

\usepackage{amsmath,amsfonts}
\usepackage{algorithmic}
\usepackage{graphicx}
\usepackage{textcomp}
\usepackage{xcolor}

\usepackage{amsmath,amsfonts}
\usepackage{algorithmic}
\usepackage{graphicx}
\usepackage{textcomp}
\usepackage{xcolor}
\usepackage{makecell}
\usepackage{mathrsfs}
\usepackage{amsthm}
\usepackage{epstopdf}
\usepackage{balance}

\usepackage{multirow}

\usepackage{epsfig,endnotes}
\usepackage{subfig}
\usepackage{grffile}
\usepackage[font=bf]{caption}
\usepackage{color, url}
\usepackage{xspace} 
\usepackage[ruled,linesnumbered]{algorithm2e}
\usepackage{epstopdf}
\usepackage{balance}
\usepackage{bm}
\usepackage{rotating}

\usepackage{enumitem}
\usepackage{makecell}
\usepackage{pdfx}

\usepackage{bm}

\newcommand{\name}{\text{StolenEncoder}}
\renewcommand{\mathbf}[1]{\bm{#1}}

\newcommand\CR[1]{{#1}}

\allowdisplaybreaks

\newcounter{subcopyrightbox@save}

\copyrightyear{2022}
\acmYear{2022}
\setcopyright{acmcopyright}
\acmConference[CCS '22] {the 2022 ACM SIGSAC Conference on Computer and Communications Security}{November 7--11, 2022}{Los Angeles, U.S.A.}
\acmBooktitle{Proceedings of the 2022 ACM SIGSAC Conference on Computer and Communications Security (CCS '22), November 7--11, 2022, Los Angeles, U.S.A.}
\acmPrice{15.00}
\acmISBN{978-1-4503-8454-4/21/11}
\acmDOI{10.1145/XXXXXXXX}

\newcommand{\myparatight}[1]{\smallskip\noindent{\bf {#1}:}~}

\graphicspath{ {figs/} }

\begin{document}

\date{}
\fancyhf{}

\title{{\name}: Stealing Pre-trained Encoders in \\Self-supervised Learning}

\author{Yupei Liu, Jinyuan Jia, Hongbin Liu, and Neil Zhenqiang Gong}
\affiliation{%
  \institution{Duke University}
}
\email{{yupei.liu, jinyuan.jia, hongbin.liu, neil.gong}@duke.edu}

\begin{abstract}
Pre-trained encoders are general-purpose feature extractors that can be used for many downstream tasks. Recent progress in self-supervised learning can pre-train highly effective encoders using a large volume of \emph{unlabeled} data, leading to the emerging \emph{encoder as a service (EaaS)}. A pre-trained encoder may be deemed confidential  because its training requires lots of data and computation resources as well as its public release may facilitate misuse of AI, e.g., for deepfakes generation.  In this paper, we propose  the first attack called \emph{{\name}} to steal pre-trained image encoders. 
We evaluate {\name} on multiple target encoders pre-trained by ourselves and three real-world target encoders including the ImageNet encoder pre-trained by Google, CLIP encoder pre-trained by OpenAI, and Clarifai's General Embedding encoder deployed as a paid EaaS. Our results show that our stolen encoders have similar functionality with the target encoders. In particular, the downstream classifiers built upon a target encoder and a stolen one have similar accuracy. Moreover, stealing a target encoder using {\name}  requires much less data and computation resources than  pre-training it from scratch. We also explore three defenses that perturb feature vectors produced by a target encoder. Our results show these defenses are not enough to mitigate {\name}.

\end{abstract}

    


\maketitle
\section{Introduction}
\begin{figure}[!t]
	 \centering
{\includegraphics[width=0.45\textwidth]{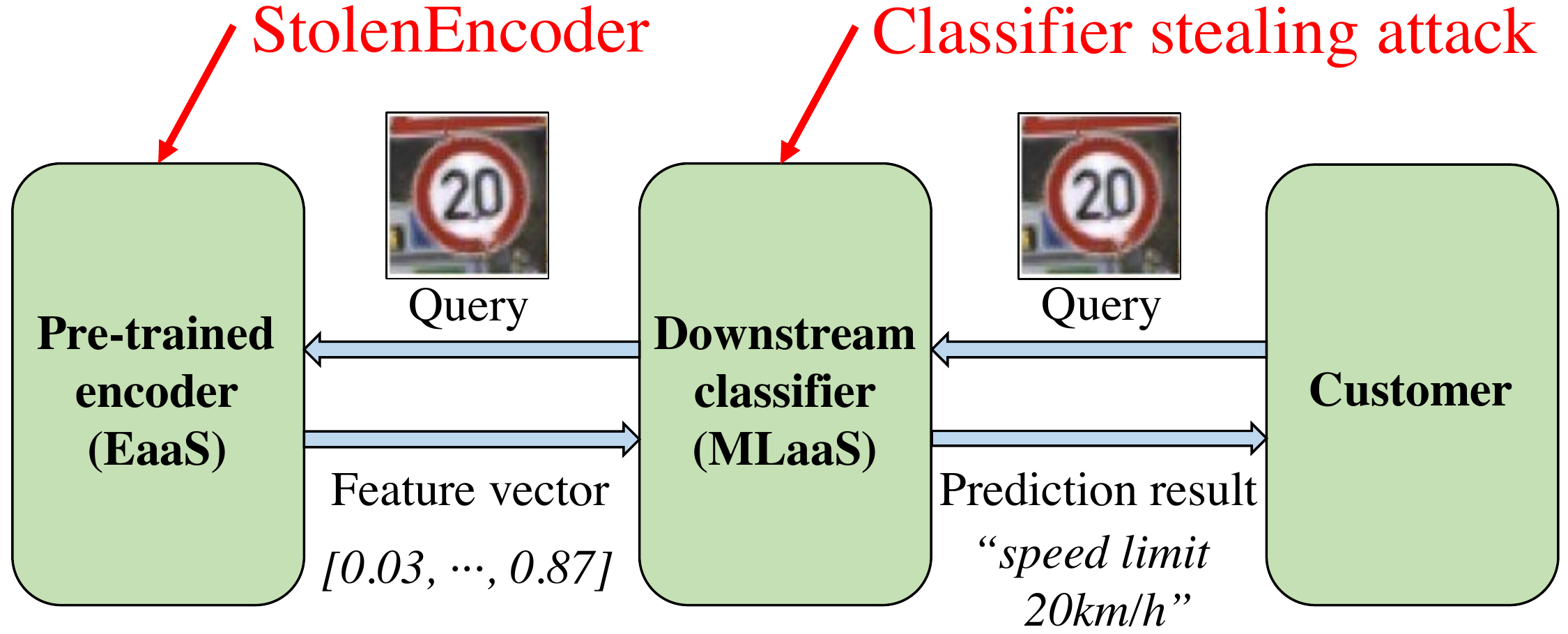}}
\caption{EaaS vs. MLaaS. } 
\label{overview}
 % \vspace{-4mm}
\end{figure}

Conventional supervised learning requires labeling a large volume of training data for each AI task.  \emph{Self-supervised learning}~\cite{he2020momentum,chen2020simple,radford2021learning} shifts the paradigm by first pre-training a general-purpose encoder using a large volume of \emph{unlabeled} data and then applying the encoder as a feature extractor for various downstream tasks with little or even no labeled training data. The success of self-supervised learning leads to an emerging cloud service, called \emph{Encoder as a Service (EaaS)}~\cite{OpenAIAPI,clarifai_image_embedding}. In particular, a powerful service provider (e.g., OpenAI, Google, and Clarifai) with sufficient data and computation resources pre-trains a general-purpose encoder and deploys it as a paid cloud service; and a customer queries the cloud service API for the feature vectors of its training/testing inputs in order to train/test  its classifiers (called \emph{downstream classifiers}). EaaS is different from conventional \emph{Machine Learning as a Service (MLaaS)} (shown in Figure~\ref{overview}), in which a (downstream) classifier is deployed as a cloud service and a customer queries the API for the prediction results of its testing inputs. 

An encoder may be kept confidential because of the large amount of data and computation resources required to pre-train it as well as its potential misuse, e.g., for deepfakes generation. For instance, OpenAI publicly released its pre-trained language encoders GPT~\cite{radford2018improving} and GPT-2~\cite{radford2019language}, but deployed its more advanced GPT-3~\cite{brown2020language} as an EaaS for both economic and ethical considerations~\cite{OpenAIAPI}, e.g., pre-trained language encoders can be misused to generate fake news~\cite{zellers2020defending}. In this work, we study attacks to steal a pre-trained encoder (called \emph{target encoder}) deployed as an EaaS, which demonstrates that deploying an encoder as a cloud service is insufficient to protect its confidentiality. An attacker can deploy the stolen encoder as its own paid EaaS or use it for its downstream tasks without querying the target EaaS.  Most existing model stealing attacks focus on stealing (downstream) classifiers~\cite{orekondy2019knockoff,chandrasekaran2020exploring,tramer2016stealing, jagielski2020high,carlini2020cryptanalytic,kariyappa2021maze, zhu2021hermes}, which are different from  stealing encoders, as illustrated in Figure~\ref{overview}. In particular, a classifier returns a label or confidence score vector, while an encoder returns a feature vector; and the loss function used to measure a label or confidence score vector is insufficient to measure a feature vector. As a result, as our experiments show, extending classifier stealing attacks to steal pre-trained encoders achieves suboptimal performance.  

\myparatight{Our work} In this work, we propose {\name}, the first attack to steal a target encoder in self-supervised learning. In particular, we focus on stealing image encoders.  {\name} aims to achieve two goals: 1) maintaining the functionality of a target encoder, and 2) the number of queries to the EaaS API is small. We consider an attacker has  a small number of unlabeled images called \emph{surrogate dataset}. The surrogate dataset does not need to follow the same distribution as the dataset used to pre-train the target encoder. 
We formulate {\name} as an optimization problem, the solution of which is a stolen encoder.

Specifically, to achieve the first goal, we first define a loss term $\mathcal{L}_1$, which is small if the stolen encoder and the target encoder output similar feature vectors for the images in the surrogate dataset. Formally, we define $\mathcal{L}_1$ as the distance (e.g., $\ell_2$ distance) between the feature vector outputted by the target encoder and that outputted by the stolen encoder for an image in the surrogate dataset on average. However,  minimizing the loss term $\mathcal{L}_1$ alone leads to a stolen encoder with suboptimal functionality because the surrogate dataset is small. To address the challenge, we propose to augment the surrogate dataset via data augmentations such as RandomHorizontalFlip, ColorJitter, and RandomGrayScale. Specifically, during the training of a stolen encoder, for each image in the surrogate dataset, we generate augmented images. Moreover, we propose another loss term $\mathcal{L}_2'$, which is small if the stolen encoder and target encoder output similar feature vectors for the augmented images. Formally, we define $\mathcal{L}_2'$ as the distance between the feature vector outputted by the target encoder and that outputted by the stolen encoder for an augmented image on average. 

However, $\mathcal{L}_2'$ requires querying the EaaS  for the feature vectors of the augmented images, which incurs a large amount of queries and violates the second goal. To  achieve the second goal, we make a key observation, i.e., a target encoder in self-supervised learning produces similar feature vectors for an image and its augmented version. Based on this observation, we approximate the loss term  $\mathcal{L}_2'$ as a loss term $\mathcal{L}_2$, which measures the distance between the feature vector outputted by the target encoder for an image and that outputted by the stolen encoder for the image's augmented version. Finally, we minimize the weighted sum of the loss terms  $\mathcal{L}_1$ and $\mathcal{L}_2$ via a stochastic gradient descent based method to obtain a stolen encoder. The number of queries to the EaaS incurred by {\name} is the size of the surrogate dataset.

We evaluate {\name} on  target encoders pre-trained by ourselves on three datasets (STL10, Food101, and CIFAR10) as well as three real-world target encoders, i.e., the ImageNet encoder pre-trained by Google~\cite{chen2020simple}, the CLIP encoder pre-trained by OpenAI~\cite{radford2021learning}, and the Clarifai General Embedding encoder deployed as a paid EaaS~\cite{clarifai_image_embedding}. Moreover, we  evaluate the functionality of an encoder using the accuracy of four downstream classifiers trained upon the encoder for four downstream datasets (MNIST, Fashion-MNIST, SVHN, and GTSRB). 
Our results show that, even if an attacker does not know the distribution of the dataset used to pre-train the target encoder, the neural network architecture of the target encoder, and the algorithm used to pre-train the target encoder, {\name} can steal an encoder with similar functionality with the target encoder using much less data and computation resources than pre-training the target encoder from scratch. For instance, the accuracy of the four downstream classifiers built upon our stolen CLIP encoder is at least 93\% of that built upon the target CLIP encoder. The target CLIP encoder was trained for 432 hours on 592 V100 GPUs using  400 million public image-text pairs collected from the Internet~\cite{radford2021learning}, while our stolen CLIP encoder was trained for only 53.9 hours on 1 Quadro-RTX-6000 GPU using 113,000 images in  STL10, which is less than 0.03\% of the size of the data used to pre-train the target CLIP encoder. Moreover, training and testing the four downstream classifiers using the Clarifai EaaS costs us \$931.6, while stealing the Clarifai encoder costs us only \$16.

We also generalize three defenses~\cite{tramer2016stealing,orekondy2019knockoff,orekondy2020prediction} against classifier stealing attacks to defend against {\name}. The key idea of these defenses is to perturb a feature vector outputted by the target encoder before returning it to a customer/attacker. The first defense (called \emph{top-$k$ features}) resets the features, whose absolute values are not the top-$k$ largest ones, to be 0. The second defense (called \emph{feature rounding}) rounds each feature in a feature vector. The third defense (called \emph{feature poisoning}) adds carefully crafted perturbation to a feature vector such that the feature vectors become a "data poisoning attack" to the training of a stolen encoder. We find that these defenses are insufficient to mitigate {\name}. In particular, feature rounding does not decrease the functionality of a stolen encoder, while top-$k$ features and feature poisoning reduce the functionality of a stolen encoder by reducing the functionality of the target encoder. 

To summarize, our work makes the following  contributions:
\begin{itemize}
    \item We propose {\name}, the first  attack to steal pre-trained encoders in self-supervised learning.
    
    \item We formulate {\name} as an optimization problem. Moreover, we propose a method based on stochastic gradient descent to solve the optimization problem. 
    
    \item We evaluate {\name} on multiple encoders trained by ourselves and three real-world image encoders. 
    
    \item We explore three defenses to mitigate {\name}. Our results highlight the needs of new mechanisms to defend against {\name}. 
\end{itemize}

\section{Preliminaries}
\label{sec:background}

Self-supervised learning~\cite{he2020momentum,chen2020simple,radford2021learning} aims to exploit the supervisory signals among unlabeled data itself to pre-train encoders, which can be used as general-purpose feature extractors for various downstream tasks. We focus on image encoders in this work. 
\emph{Contrastive learning}~\cite{he2020momentum,chen2020simple,radford2021learning} is a representative family of self-supervised learning techniques that pre-train image encoders. Therefore, we focus on contrastive learning in the following. Contrastive learning pre-trains an image encoder using a large volume of unlabeled images or image-text pairs. We call such  unlabeled data \emph{pre-training dataset}. 
Next, we first discuss how to pre-train an image encoder and then discuss how to apply it to train downstream classifiers.

\subsection{Pre-training an Encoder}
One major component of contrastive learning is the stochastic data augmentation module, which consists of a sequence of random augmentation operations, e.g., randomly horizontally flipping an image (i.e., RandomHorizontalFlip), randomly changing the brightness, contrast and saturation of an image (i.e., ColorJitter), as well as randomly converting an image to grayscale (e.g., RandomGrayScale). Given an input image, the data augmentation module generates two stochastic augmented views called \emph{positive pair}. Two augmented views from different input images are called \emph{negative pair}. Generally speaking, the core idea of contrastive learning is to pre-train an image encoder that outputs similar feature vectors for positive pairs but dissimilar feature vectors for negative pairs. Next, we discuss two state-of-the-art contrastive learning algorithms, SimCLR~\cite{chen2020simple} and MoCo~\cite{he2020momentum}, in more details. Both algorithms use unlabeled images to pre-train an image encoder.

\myparatight{SimCLR~\cite{chen2020simple}} Besides the data augmentation module,  SimCLR  contains two other important modules: image encoder $f$ and projection head $h$. Given an input image $\boldsymbol{x}$, the image encoder $f$ outputs the feature vector $f(\boldsymbol{x})$. The projection head $h$ maps the feature vectors to calculate contrastive loss. The projection head $h$ is a multilayer perceptron with one hidden layer. Given a randomly sampled minibatch of $N$ images $\{\boldsymbol{x_1}, \boldsymbol{x_2},\cdots,\boldsymbol{x_{N}} \}$, SimCLR generates two augmented views from each image, resulting in $2 N$ augmented images: $\{\widetilde{\boldsymbol{x}}_1, \widetilde{\boldsymbol{x}}_2,\cdots,\widetilde{\boldsymbol{x}}_{2N} \}$. Given a positive pair $\{\widetilde{\boldsymbol{x}}_i, \widetilde{\boldsymbol{x}}_j\}$, $\widetilde{\boldsymbol{x}}_i$ and the remaining $2(N-1)$ images form negative pairs. SimCLR defines the contrastive loss for a positive pair  $\{\widetilde{\boldsymbol{x}}_i, \widetilde{\boldsymbol{x}}_j\}$ as follows:
\begin{align}
\ell_{i, j}=-\log \frac{\exp \left(\operatorname{sim}\left(h(f(\widetilde{\boldsymbol{x}}_i)), h(f(\widetilde{\boldsymbol{x}}_j))\right) / \tau\right)}{\sum_{k=1,k \neq i}^{2 N} \exp \left(\operatorname{sim}\left(h(f(\widetilde{\boldsymbol{x}}_i)), h(f(\widetilde{\boldsymbol{x}}_k))\right) / \tau\right)},
\end{align}
where $\operatorname{sim}$ denotes the cosine similarity and $\tau$ denotes a temperature hyperparameter. The overall contrastive loss is the sum of $\ell_{i, j}$ over all positive pairs. SimCLR pre-trains an encoder $f$ together with the projection head $h$ via minimizing the overall contrastive loss.

\myparatight{MoCo~\cite{he2020momentum}} Besides the stochastic data augmentation module, the MoCo framework consists of three major components: query image encoder $f$, momentum image encoder $f_m$, and dictionary $\mathcal{D}$. Both query image encoder $f$ and momentum image encoder $f_m$  output feature vectors for input images and they have the same architecture. Given input images' augmented views, the feature vectors outputted by the momentum image encoder $f_m$ are called key vectors. The dictionary $\mathcal{D}$ consists of a queue of key vectors that are created in preceding several minibatches. The dictionay $\mathcal{D}$ is updated iteratively by adding key vectors in the current minibatch and removing the key vectors from the oldest minibatch. MoCo designs the momentum image encoder $f_m$ to be updated significantly slower than the query image encoder $f$ so that key vectors maintain representations’ consistency in dictionary $\mathcal{D}$. Similar to SimCLR~\cite{chen2020simple}, given a minibatch of $N$ images $\{\boldsymbol{x_1}, \boldsymbol{x_2},\cdots,\boldsymbol{x_{N}} \}$, MoCo also generates two augmented views from each image to obtain $2 N$ augmented images: $\{\widetilde{\boldsymbol{x}}_1, \widetilde{\boldsymbol{x}}_2,\cdots,\widetilde{\boldsymbol{x}}_{2N} \}$. For each positive pair $\{\widetilde{\boldsymbol{x}}_i, \widetilde{\boldsymbol{x}}_j\}$ generated from the same image $\boldsymbol{x}$, $\widetilde{\boldsymbol{x}}_i$ denotes query and $\widetilde{\boldsymbol{x}}_j$ denotes key. They are respectively inputted to the query image encoder $f$ and momentum image encoder $f_m$ to obtain query feature vector $f(\boldsymbol{x}_i)$ and key vector $f_m(\boldsymbol{x}_j)$. Then the key vector $f_m(\boldsymbol{x}_j)$ is enqueued to the dictionary $\mathcal{D}$. MoCo defines the contrastive loss for a positive pair of query $\widetilde{\boldsymbol{x}}_i$ and key $\widetilde{\boldsymbol{x}}_j$ as follows:
\begin{align}
\ell_{i, j}=-\log \frac{\exp \left(\operatorname{sim}\left(f(\widetilde{\boldsymbol{x}}_i), f_m(\widetilde{\boldsymbol{x}}_j)\right) / \tau\right)}{\sum_{\boldsymbol{k}\in \mathcal{D}} \exp \left(\operatorname{sim}\left(f(\widetilde{\boldsymbol{x}}_i), \boldsymbol{k}\right) / \tau\right)},
\end{align}
where $\operatorname{sim}$ denotes the cosine similarity and $\tau$ denotes a temperature hyperparameter. The overall contrastive loss computes the sum of $\ell_{i, j}$ over all $N$ positive pairs. MoCo pre-trains an encoder $f$ via minimizing the overall contrastive loss.

\subsection{Training/Testing Downstream Classifiers}

The image encoder can be used to output feature vectors for various downstream tasks, which we consider to be image classification in this work. We can use the pre-trained image encoder to extract feature vectors for training images, and then train a downstream classifier on the feature vectors and corresponding labels following the supervised learning paradigm. Given a testing image, we use the image encoder to output its feature vector and then use the downstream classifier to predict its label. We call the dataset used to train and test a downstream classifier \emph{downstream dataset}.
\section{Threat Model}

We consider two parties: \emph{EaaS provider} and \emph{attacker}. In particular, an EaaS  provider could be a resourceful  entity such as Google, OpenAI, and Clarifai.  The EaaS provider uses contrastive learning to pre-train an encoder (called \emph{target encoder}) on a pre-training dataset and deploys it as a paid EaaS. 
 A customer (e.g., an attacker) can query the EaaS API to obtain the feature vectors for its input images. In particular,  a customer sends an input image to the EaaS  API, which uses the target encoder to compute the feature vector for the image and  returns it to the customer. The EaaS  may charge a customer based on the number of queries sent by the customer.  Next, we  discuss our threat model with respect to the attacker's goals, background knowledge, and capabilities. 

\myparatight{Attacker's goals}  An attacker aims to achieve two goals:
\begin{itemize}
    \item {\bf Goal I: Maintaining the functionality of the target encoder.} In the first goal, the attacker aims to steal the target encoder such that the stolen encoder maintains the functionality of the target encoder. In particular, the downstream classifiers built upon the stolen encoder should be as accurate as those built upon the target encoder for different downstream tasks. 
    
    \item {\bf Goal II: A small query cost.}  The number of queries to the EaaS API represents the economic cost for the attacker. 
    Therefore, in the second goal, the attacker aims to steal the target encoder using a small number of queries.  
\end{itemize}

\myparatight{Attacker's background knowledge} We consider the attacker's background knowledge along two dimensions:
\begin{itemize}
    \item {\bf Surrogate dataset.} We assume the attacker has a small amount of unlabeled images (called \emph{surrogate dataset}). We consider three scenarios for the surrogate dataset depending on the attacker's background knowledge on the pre-training dataset of the target encoder. In the first scenario, the surrogate dataset is a subset of the pre-training dataset. For instance, when the EaaS provider collects public data on the Internet as the pre-training dataset, the attacker can also collect some public data from the Internet as the surrogate dataset. In the second scenario, the surrogate dataset has the same distribution as the pre-training dataset, but does not have overlap with the pre-training dataset. In the third scenario,  the surrogate dataset has a different distribution from the pre-training dataset, which means that the attacker does not know the distribution of the pre-training dataset, e.g., when the pre-training dataset is the EaaS provider's private data. As our experiments show, our attack achieves similar effectiveness in the three scenarios and our stolen encoder outperforms an encoder pre-trained on a surrogate dataset locally.

    \item {\bf Encoder architecture.} This background knowledge characterizes whether the attacker knows the neural network architecture of the target encoder. If the attacker knows the target encoder's architecture, e.g., the EaaS provider makes it public to increase transparency and trust, the attacker can use the same architecture for its stolen encoder. When the attacker does not know the target encoder's architecture, the attacker can select an expressive/powerful architecture. For instance, the attacker can use ResNet-34~\cite{he2016deep} as its stolen encoder architecture. The attacker selects an expressive architecture so the stolen encoder is more likely to maintain the functionality of the target encoder.  
    
\end{itemize}

We note that the attacker does not need to know the contrastive learning algorithm used to pre-train the target encoder. This is because our  attack does not use a contrastive learning algorithm to train the stolen encoder, as we will discuss in the next section.

\myparatight{Attacker's capabilities} An attacker can  query the EaaS API to obtain the feature vectors of its input images.

\section{StolenEncoder}
\label{sec:attacks}

\subsection{Overview}

Our key idea is to formulate our encoder stealing attack as an optimization problem and then leverage the standard stochastic gradient descent to solve the optimization problem, which produces the stolen encoder. Recall that our attack aims to achieve two goals. To achieve Goal I, we require that the stolen encoder and the target encoder produce similar feature vectors for the images in the surrogate dataset. Moreover, since the surrogate dataset is small, we  use data augmentations to augment each image in the surrogate dataset, and we require that the stolen encoder and the target encoder also produce similar feature vectors for the augmented images. However, such requirement incurs a large amount of queries to the EaaS API as the attacker needs to obtain the feature vectors of the augmented images produced by the target encoder, which does not satisfy Goal II. To address the challenge, our key observation is that a target encoder pre-trained by contrastive learning produces similar feature vectors for an image and its augmented versions. Therefore, to achieve Goal II, we propose to approximate the feature vector of an augmented image as that of the original image produced by the target encoder. Such approximation eliminates the needs of querying the EaaS API for the feature vectors of the augmented images, which reduces the query cost substantially (i.e., achieves Goal II) while still achieving Goal I. 

\subsection{Formulating an Optimization Problem}
\label{formulate_optimization_problem}

We use $f_t$ to denote the target encoder and use $f_s$ to denote the stolen encoder. We use $\mathcal{D}$ to denote the surrogate dataset of  the attacker. $\mathbf{x}$ is an input image, and $f_t(\mathbf{x})$ (or $f_s(\mathbf{x})$) is the feature vector produced by $f_t$ (or $f_s$) for $\mathbf{x}$. Next, we  formally quantify the attacker's two goals, based on which we formulate {\name} as an optimization problem.

\myparatight{Achieving Goal I} Goal I aims to maintain the functionality of the target encoder. The functionality of an encoder is often measured by the accuracy of the downstream classifiers built upon it. However, it may be unknown which downstream classifiers will be built upon the stolen encoder at the time of stealing. As a result, it is challenging to directly quantify Goal I using accuracy of the downstream classifiers. To address the challenge, we propose to quantify Goal I using the outputs of the stolen encoder itself. Intuitively, if the stolen encoder and the target encoder produce similar feature vectors for any input, then the downstream classifiers built upon them would have similar accuracy. Based on this intuition, we quantify Goal I using the feature vectors produced by the stolen encoder and the target encoder. 

Specifically, the stolen encoder and target encoder should produce similar feature vectors for the images in the attacker's surrogate dataset, which we quantify using the following loss  $\mathcal{L}_1$:  
\begin{align}
\label{utility_loss_1}
\mathcal{L}_1 = \frac{1}{|\mathcal{D}|}\cdot \sum_{\mathbf{x}\in \mathcal{D}} d(f_t(\mathbf{x}),f_s(\mathbf{x})),
\end{align}
where $f_s$ is the stolen encoder, $\mathbf{x}$ is an input image in the surrogate dataset $\mathcal{D}$, and $d$ is a distance metric (e.g., $\ell_2$-distance)  to measure the distance between the feature vectors $f_t(\mathbf{x})$ and $f_s(\mathbf{x})$, which are produced by the target encoder and stolen encoder for $\mathbf{x}$, respectively.  $\mathcal{L}_1$ is small if  the stolen encoder $f_s$ and the target encoder $f_t$ produce similar feature vectors for an input.

In our threat model, we assume the attacker's surrogate dataset is small because the attacker is often a less resourceful entity who cannot pre-train its own encoder. Therefore, minimizing the loss term $\mathcal{L}_1$ alone with a small surrogate dataset may lead to a stolen encoder with suboptimal functionality, as shown in our experiments. 
To address the challenge, inspired by contrastive learning, we use data augmentation to augment the surrogate dataset and require the stolen encoder  to produce similar feature vectors for the augmented images with the target encoder. Specifically, for each image in the surrogate dataset, we create an augmented image via applying a series of random data augmentation operations (e.g., RandomHorizontalFlip, ColorJitter, RandomGrayScale). For simplicity, we use $\mathcal{A}$ to denote the composition of a series of data augmentation operations and use $\mathcal{A}(\mathbf{x})$ to denote an augmented image created from $\mathbf{x}$. 
Given those augmented images, we require the stolen encoder to produce similar feature vectors for them  with the target encoder. Formally, we  define the following loss term $\mathcal{L}_2^{\prime}$:
\begin{align}
\label{utility_loss_2_temp}
    \mathcal{L}_2^{\prime} = \frac{1}{|\mathcal{D}|} \cdot \sum_{\mathbf{x}\in \mathcal{D}} d(f_t(\mathcal{A}(\mathbf{x})),f_s(\mathcal{A}(\mathbf{x}))),
\end{align}
where $\mathcal{A}(\mathbf{x})$ is an augmented image created from $\mathbf{x}$ in $\mathcal{D}$. Given the two loss terms defined in Equation~(\ref{utility_loss_1}) and~(\ref{utility_loss_2_temp}), we can achieve Goal I via training the stolen encoder $f_s$ by solving the following optimization problem:
\begin{align}
\label{temple_final_loss}
   \min_{f_s} \mathcal{L}' = \mathcal{L}_1 + \lambda \cdot \mathcal{L}_2^\prime,
\end{align}
where $\lambda$ is a hyperparameter that balances the two loss terms.

\myparatight{Achieving Goal II} Suppose the attacker solves the optimization problem in Equation~(\ref{temple_final_loss}) using the standard Stochastic Gradient Descent (SGD). Specifically, the attacker randomly initializes the stolen encoder. The attacker repeatedly calculates the gradient of the loss function $\mathcal{L}'$ with respect to the stolen encoder using a mini-batch of images in the surrogate dataset, and the attacker moves the stolen encoder towards the inverse of the gradient with a small step. The attacker can repeat this process for $e$ epochs. 

In the above optimization process, the attacker needs to query the target encoder once for the feature vector of each image in the surrogate dataset, which can be saved locally and used to calculate the loss term $\mathcal{L}_1$ during the optimization process. In other words, the loss term $\mathcal{L}_1$ incurs $|\mathcal{D}|$ queries.  Moreover, the attacker also  needs to  query the target encoder for the feature vectors of the augmented images. This results in a large query cost, especially when a different augmented image is used in each epoch of the optimization process to enhance the functionality of the stolen encoder.  Specifically, the attacker needs to query the target encoder   $e\cdot |\mathcal{D}|$ times to obtain the feature vectors of the augmented images in calculating the loss term $\mathcal{L}_2'$ during the optimization process. Therefore, the total query cost is  $(e+1)\cdot |\mathcal{D}|$.

The query cost to calculate the loss term $\mathcal{L}_1$ is already minimal, i.e., one query  per image in the surrogate dataset. Therefore, we aim to reduce the query cost of calculating the loss term $\mathcal{L}_2'$. Our key observation is that the target encoder pre-trained by contrastive learning produces similar feature vectors for an image and its augmented version. In other words, we have $f_t(\mathcal{A}(\mathbf{x}))\approx f_t(\mathbf{x})$. Based on this observation, we propose to use $f_t(\mathbf{x})$ to approximate $f_t(\mathcal{A}(\mathbf{x}))$. Formally, 
we define the following loss term $\mathcal{L}_2$:
\begin{align}
\label{utility_loss_2}
    \mathcal{L}_2 = \frac{1}{|\mathcal{D}|} \cdot \sum_{\mathbf{x}\in \mathcal{D}} d(f_t(\mathbf{x}),f_s(\mathcal{A}(\mathbf{x}))).
\end{align}

We use $\mathcal{L}_2$ to approximate $\mathcal{L}_2^{\prime}$. In other words, we transform the optimization problem in Equation~(\ref{temple_final_loss}) as the following:
\begin{align}
    \label{final_loss}
   \min_{f_s} \mathcal{L} = \mathcal{L}_1 + \lambda \cdot \mathcal{L}_2,
\end{align}
where $\lambda$ is a hyperparameter. As our experiments show, the stolen encoders obtained by solving Equation~(\ref{temple_final_loss}) and Equation~(\ref{final_loss}) achieve similar functionality. However, the loss term $\mathcal{L}_2$ does not incur extra queries to the target encoder, since the feature vectors $f_t(\mathbf{x})$ are already obtained and saved locally for calculating $\mathcal{L}_1$.  Thus,  the total query cost of solving Equation~(\ref{final_loss}) is only $|\mathcal{D}|$.

\begin{algorithm}[tb]
   \caption{\name}
   \label{alg:example}
\begin{algorithmic}[1]
   \STATE {\bfseries Input:}\textsl{API} (EaaS API of target encoder $f_t$),  $\mathcal{D}$ (surrogate dataset), $d$ (distance metric), $e$ (number of epochs), $lr$ (learning rate), $\lambda$ (hyperparameter), $\mathcal{A}$ (composition of data augmentation operations), $S$ (batch size).
   \STATE {\bfseries Output:}  $\Theta$ (parameters of stolen encoder $f_s$). \\
   \STATE $\backslash\backslash$ Query the EaaS API for feature vectors\\
   \STATE $\mathbf{v}(\mathbf{x}) \gets \textsl{API}(\mathbf{x}), \mathbf{x}\in \mathcal{D}$ \label{alg:query_API} \\
     \STATE $\backslash\backslash$ Randomly initialize the stolen encoder parameters $\Theta$  \\
   \STATE $\Theta \gets \textsl{RandomIni}()$\label{alg:random_ini} \\
   \FOR{$j=1,2,\cdots,e$}
   \FOR{$i=1,2,\cdots, \lfloor |\mathcal{D}|/S \rfloor$}
   \STATE $\mathcal{MB}\gets \textsl{MiniBatch}(\mathcal{D})$ \label{alg:mini-batch}\\
   \STATE $ \mathcal{L}_1 \gets \frac{1}{|\mathcal{MB}|}\cdot \sum_{\mathbf{x}\in \mathcal{MB}} d(\mathbf{v}(\mathbf{x}),f_s(\mathbf{x}))$ \label{alg:compute_L_1} \\
   \STATE  $\mathcal{L}_2 \gets \frac{1}{|\mathcal{MB}|} \cdot \sum_{\mathbf{x}\in \mathcal{MB}} d(\mathbf{v}(\mathbf{x}),f_s(\mathcal{A}(\mathbf{x})))$ \label{alg:compute_L_2} \\
   \STATE $\Theta \gets \Theta - lr \cdot \frac{\partial (\mathcal{L}_1+\lambda \cdot \mathcal{L}_2)}{\partial \Theta}$ \label{alg:update_parameters}
   \ENDFOR
   \ENDFOR
   \STATE \textbf{return} $\Theta$
\end{algorithmic}
\end{algorithm}

\subsection{Solving the Optimization Problem}
The solution to the optimization problem in Equation~(\ref{final_loss}) is our stolen encoder. Algorithm~\ref{alg:example} shows the algorithm of $\name$. The attacker first queries the EaaS API for the feature vector $f_t(\mathbf{x})$ of each image $\mathbf{x}$ in its surrogate dataset (Line~\ref{alg:query_API} in Algorithm~\ref{alg:example}). The function \textsl{RandomIni} in Line~\ref{alg:random_ini}  randomly initializes the stolen encoder $f_s$. Then, the attacker uses the standard SGD to solve the optimization problem. In each epoch, 
the attacker processes the surrogate dataset mini-batch by mini-batch. For each mini-batch $\mathcal{MB}$ of images in $\mathcal{D}$, the attacker uses $\mathcal{MB}$ to respectively compute the two loss terms $\mathcal{L}_1$ and $\mathcal{L}_2$ in Line~\ref{alg:compute_L_1} and~\ref{alg:compute_L_2}, and finally uses gradient descent to update parameters $\Theta$ of the stolen encoder in Line~\ref{alg:update_parameters}. We note that the data augmentation $\mathcal{A}(\mathbf{x})$ generates a random augmented version of $\mathbf{x}$ in each epoch.

\section{Evaluation}
\label{sec:exp}
\subsection{Experimental Setup}

\subsubsection{Pre-training Datasets and Target Encoders} We use STL10~\cite{coates2011analysis},  Food101~\cite{bossard14}, and CIFAR10~\cite{krizhevsky2009learning} as the pre-training datasets. STL10 dataset includes 5,000 labeled training images, 8,000 labeled testing images, and 100,000 unlabeled images. Food101 dataset consists of 90,900 training images and 10,100 testing images. CIFAR10 dataset includes 50,000 labeled training images and 10,000 labeled testing images.  
Table~\ref{dataset_table} summarizes all our datasets. 

We pre-train a target encoder on each pre-training dataset. In particular, for CIFAR10 and Food101, we use their training images  to pre-train target encoders. For STL10, we use both the training images and the unlabeled images to pre-train a target encoder.  We do not use the testing images of the pre-training datasets, which we reserve to evaluate  {\name} in the scenario where the attacker's surrogate dataset  follows the same distribution as the pre-training dataset but does not have overlaps with it. By default, we adopt SimCLR~\cite{chen2020simple} as the contrastive learning algorithm to pre-train target encoders. We train a target encoder for 1,000 epochs, where the batch size is 256, the optimizer is Adam, and the initial learning rate is 0.001. We use data augmentation operations including RandomResizedCrop,  RandomHorizontalFlip,  ColorJitter, and RandomGrayScale in SimCLR. The architecture of the target encoder is ResNet18~\cite{he2016deep}. We use the public  code  of SimCLR~\cite{simclr_url}. 

\begin{table}[tp]\renewcommand{\arraystretch}{1.2} 
\fontsize{8}{8}\selectfont
\centering
	\caption{Dataset summary.}
	\setlength{\tabcolsep}{1mm}
	{
	\begin{tabular}{|c|c|c|c|}
		\hline
	 & \makecell{Dataset} & \makecell{\#Training \\Examples} & \makecell{\#Testing \\Examples} \\ \hline \hline
	\multirow{3}{*}{Pre-training Dataset}
	& STL10        & 105,000 & 8,000  \\ \cline{2-4}  
	& Food101      & 90,900 & 10,100  \\ \cline{2-4} 
	& CIFAR10 & 50,000 & 10,000  \\ \cline{1-4}  \hline \hline
	\multirow{4}{*}{Downstream Dataset}
	& MNIST        & 60,000 & 10,000  \\ \cline{2-4}  
	& FashionMNIST & 60,000 & 10,000  \\ \cline{2-4}  
	& SVHN         & 73,257 & 26,032  \\ \cline{2-4}  
		& GTSRB        & 39,209 & 12,630  \\ \cline{1-4}   \hline \hline
	Surrogate Dataset	& ImageNet      &--& --  \\ \hline
	\end{tabular}
	}
	\label{dataset_table}
	\vspace{-4mm}
\end{table}

\subsubsection{Attack Settings} {\name} has the following parameters: surrogate dataset $\mathcal{D}$, architecture of the stolen encoder, composition of data augmentation operations $\mathcal{A}$, distance metric $d$, hyperparameter $\lambda$, number of epochs $e$, learning rate $lr$, and batch size $S$. We adopt the following default settings. For surrogate dataset, we assume the attacker does not know the distribution of the pre-training dataset. In particular, we sample the surrogate dataset from the ImageNet dataset~\cite{deng2009imagenet}. Moreover, the size of the surrogate dataset is 5\% of the pre-training dataset size, i.e.,  the sizes of the surrogate dataset when stealing the target encoders pre-trained on STL10, Food101, and CIFAR10 are  5,250, 4,545, and 2,500, respectively.

We assume the attacker does not know the architecture of the target encoder. In particular, we adopt ResNet34 as the  architecture of the stolen encoder.
$\mathcal{A}$ is a composition of data augmentation operations including   RandomHorizontalFlip,  ColorJitter, and RandomGrayScale. We note that these data augmentations are not the same as those used to pre-train the target encoder. Moreover, our results on CLIP in Section~\ref{sec:casestudy} show that even if the data augmentation operations $\mathcal{A}$ used to train the stolen encoder have no overlaps with those used to pre-train the target encoder, {\name} is still effective.     

We adopt $\ell_2$ distance as the distance metric $d$. We set $\lambda=20$,  $e=100$, $lr=0.001$, and $S=64$. We will study the impact of each parameter on $\name$. Moreover, we set all other parameters to the default values when studying the impact of one parameter. All our experiments are performed on 1 Quadro-RTX-6000 GPU.

\subsubsection{Downstream Datasets and Downstream Classifiers} We use MNIST~\cite{MNIST}, FashionMNIST~\cite{FashionMnist},  SVHN~\cite{netzer2011reading},  and GTSRB~\cite{stallkamp2012man} as the four downstream datasets. MNIST dataset  includes 60,000 training images and 10,000 testing images from 10 classes. 
FashionMNIST consists of 60,000 training images and 10,000 testing images of fashion products from 10 categories. 
SVHN dataset consists of images obtained from house numbers in Google Street View images. The dataset includes 73,257 training images and 26,032 testing images from 10 classes. GTSRB dataset is a traffic sign recognition dataset, which includes 39,209 training images and 12,630 testing images from 43 classes. 
We use GTSRB as the default downstream dataset in our evaluation as it is the most challenging one. Moreover, we resize all images in the pre-training datasets and downstream datasets to be $32 \times 32$. Note that each image in MNIST and FashionMNIST only has one channel, but all other images have three channels. We extend the images in both datasets to three channels. In particular, following previous work~\cite{radford2021learning}, we extend one channel to three channels by setting the other two channels to be the same as the existing channel.

For each downstream dataset, we use its training dataset to train a downstream classifier via treating the target or stolen encoder as a feature extractor. We adopt a  fully connected neural network with two hidden layers as a downstream classifier. The number of neurons in the two hidden layers are respectively 512 and 256. Morover, we adopt ReLU as the activation function for the input and hidden layers and adopt Softmax as the activation function for the output layer. By default, we use the Adam optimizer with an initial learning rate 0.0001 and batch size  256 to train a downstream classifier for 500 epochs. We will also explore the impact of the parameters in training the downstream classifiers on {\name}.

\subsubsection{Compared Methods}\label{comparision_methods} We compare {\name} with the following baselines or variants of {\name}. 

    \myparatight{Pre-training an encoder using surrogate dataset (Pre-training-encoder)} As an attacker has a surrogate dataset, the attacker can pre-train an encoder on its surrogate dataset without querying the target encoder at all. In particular, in this  method, we use SimCLR to pre-train an encoder  on the surrogate dataset, where the parameter settings of SimCLR are the same as those used to pre-train the target encoders.

  \myparatight{{\name} without augmented images ({\name}-w/o-aug)} Recall that {\name} uses augmented images of the surrogate dataset to enhance the functionality of the stolen encoder. To demonstrate such augmented images do enhance functionality of the stolen encoder, we evaluate a variant of {\name} that does not use augmented images. That is, this variant only uses the loss term $\mathcal{L}_1$  to train the stolen encoder, which is equivalent to setting $\lambda=0$ in {\name}. 
    
    We note that {\name}-w/o-aug can also be viewed as generalizing classifier stealing attack to steal encoder. In particular, attacks~\cite{orekondy2019knockoff,chandrasekaran2020exploring,tramer2016stealing,jagielski2020high} to steal a complex neural network classifier essentially query the classifier and treat the query results (i.e., labels or confidence score vectors) as labels of the queries to train a stolen classifier.  {\name}-w/o-aug treats the query results (i.e., feature vectors outputted by the target encoder) as ``labels'' of the queries to train a stolen encoder.

  \myparatight{{\name} with high query cost ({\name}-query-aug)} Recall that we use $\mathcal{L}_2$ to approximate $\mathcal{L}_2^{\prime}$ to reduce the query cost (please refer to Section~\ref{formulate_optimization_problem} for details). To demonstrate such approximation does not sacrifice functionality of the stolen encoder, we evaluate a variant of {\name}  that uses $\mathcal{L}_2^{\prime}$ instead of $\mathcal{L}_2$. That is, in this variant, the attacker queries the cloud service API for the feature vectors of the augmented images.

\begin{figure*}[!t]
	 \centering
\subfloat[STL10]{\includegraphics[width=0.32\textwidth]{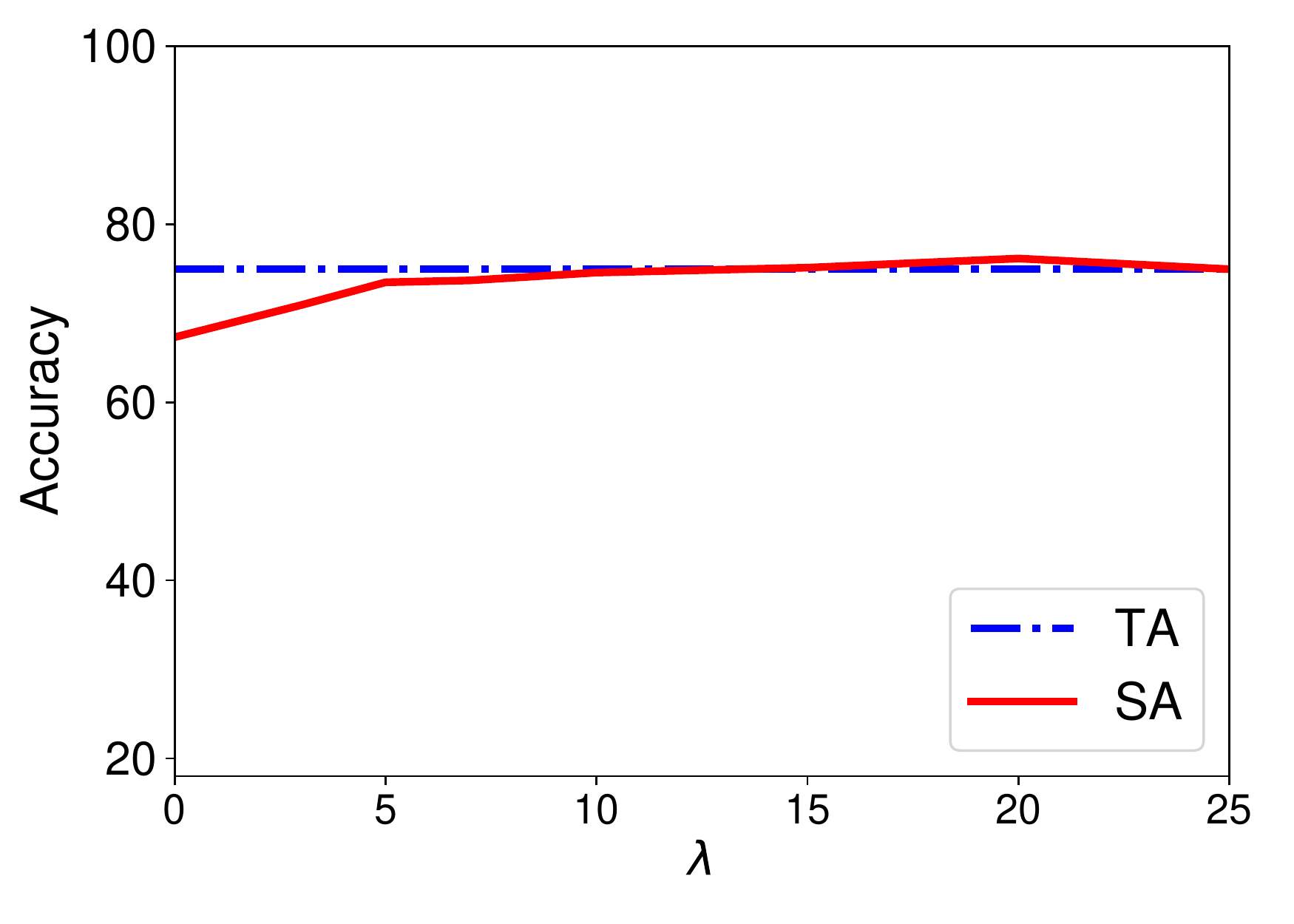}}
\subfloat[Food101]{\includegraphics[width=0.32\textwidth]{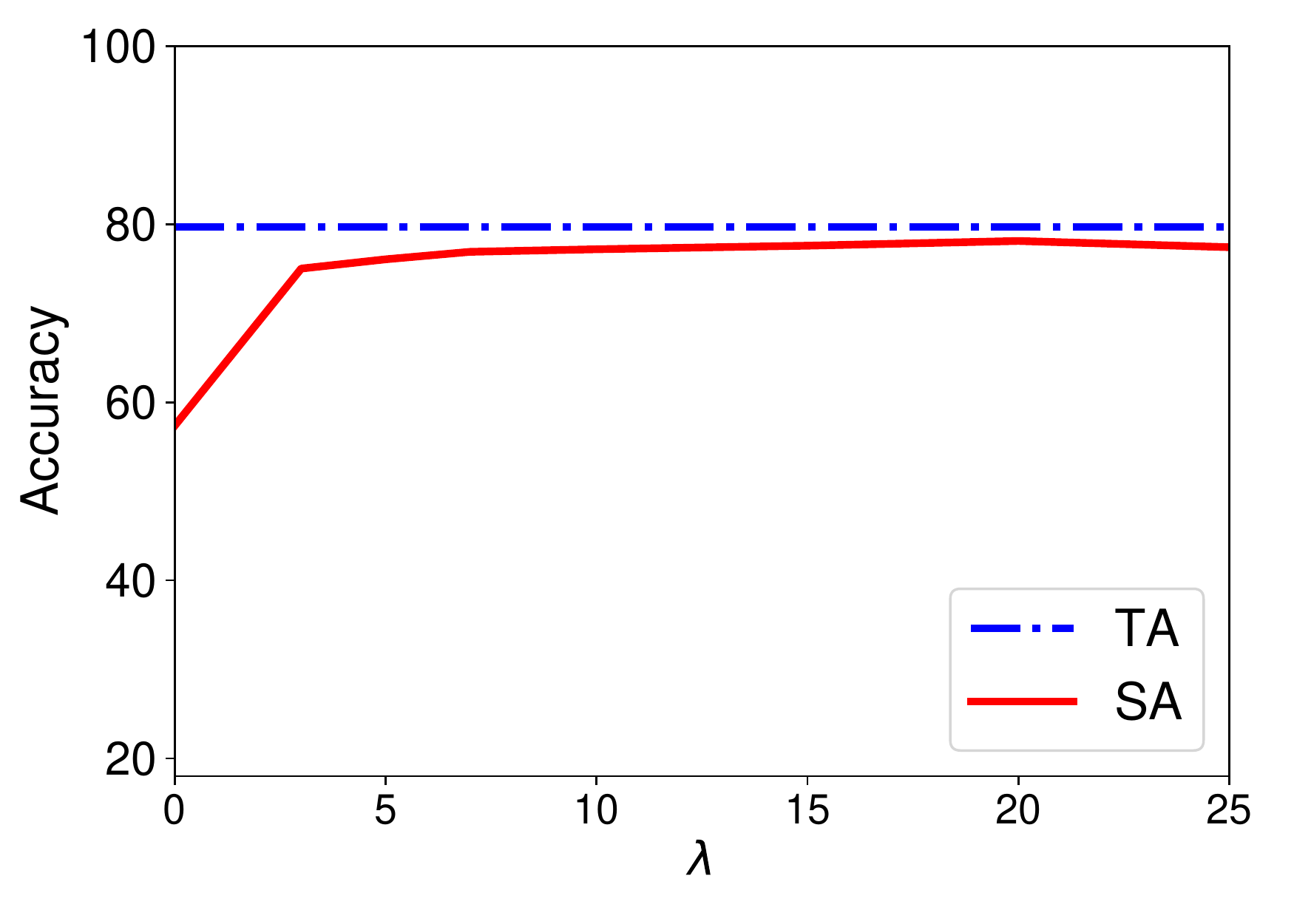}}
\subfloat[CIFAR10]{\includegraphics[width=0.32\textwidth]{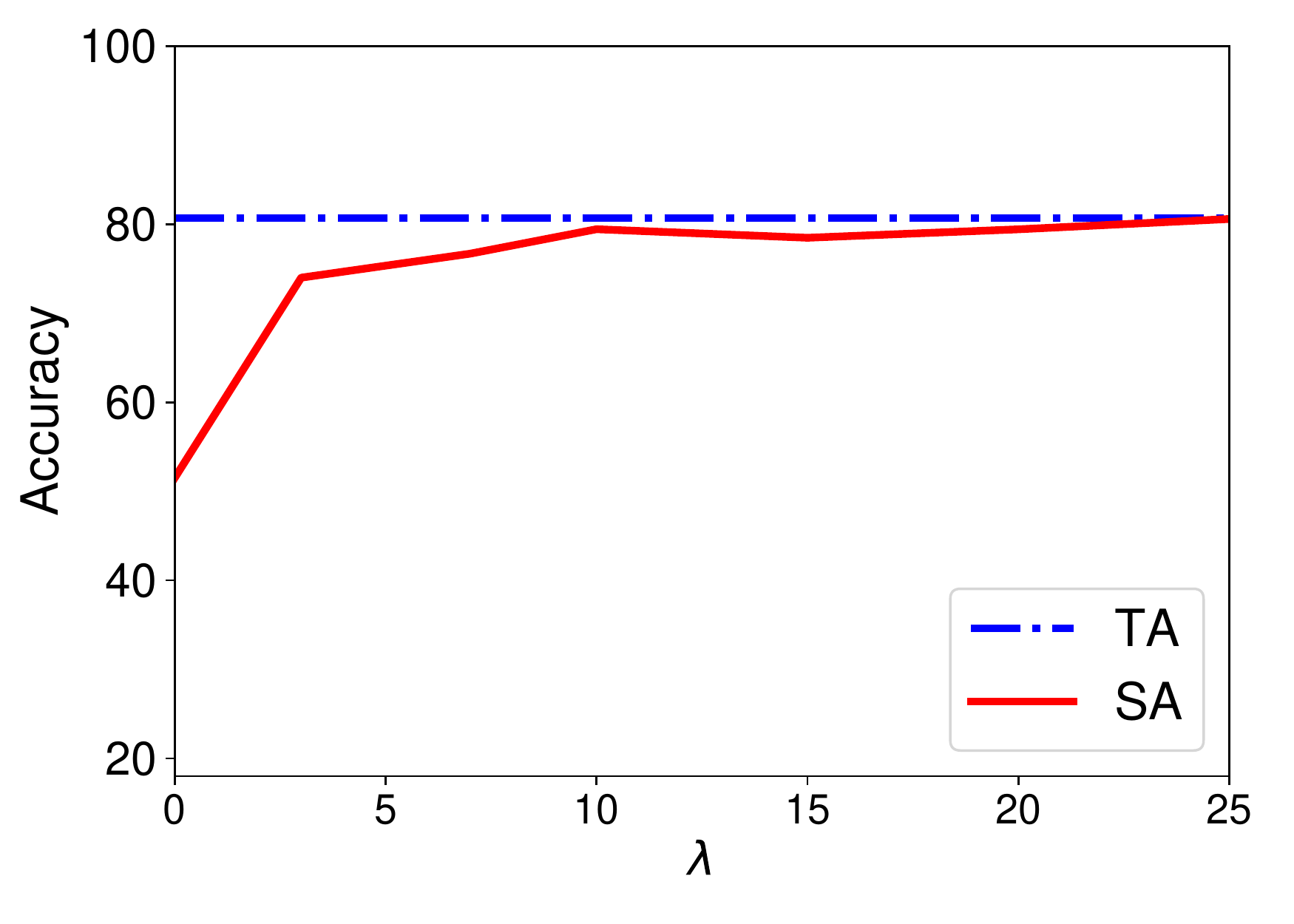}}
\vspace{-2mm}
\caption{Impact of $\lambda$ on SA of {\name} for the three pre-training datasets. } 
\label{impact_of_lambda}
\vspace{-2mm}
\end{figure*}

\begin{table}[tp]\renewcommand{\arraystretch}{1.2} 
 	\fontsize{8}{8}\selectfont
	\centering
	\caption{{\name} is effective. }
	\setlength{\tabcolsep}{1mm}
	{
	\begin{tabular}{|c|c|c|c|c|}
		\hline
	\makecell{Pre-training \\Dataset} & \makecell{Downstream \\Dataset} & \makecell{TA (\%)} & \makecell{SA (\%)}&  $\frac{\text{SA}}{\text{TA}}\times 100\%$  \\ \hline \hline
	\multirow{4}{*}{STL10}
	& MNIST        & 97.63 & 97.74 & 100 \\ \cline{2-5}  
	& FashionMNIST & 89.98 & 90.35  & 100\\ \cline{2-5}  
	& SVHN         & 56.97 & 73.50  &129 \\ \cline{2-5} 
		& GTSRB        & 74.97 & 74.57  & 99 \\
	\hline \hline
	\multirow{4}{*}{Food101}
	& MNIST        & 97.92 & 97.97  &100 \\ \cline{2-5}  
	& FashionMNIST & 89.88 & 90.63 & 101  \\ \cline{2-5}  
	& SVHN         & 62.22 & 77.50  & 125 \\ \cline{2-5}  	
	& GTSRB        & 79.69 & 78.12 & 98  \\   
\hline  \hline
	\multirow{4}{*}{CIFAR10}
	& MNIST        & 97.90 & 97.95  & 100\\ \cline{2-5}  
	& FashionMNIST & 89.44 & 90.34  &101 \\ \cline{2-5}  
	& SVHN         & 59.63 & 77.04  &129 \\ \cline{2-5}  
		& GTSRB        & 80.67 & 79.43  &98 \\ \hline 

	\end{tabular}
	}
	\label{default_table}
 	\vspace{-2mm}
\end{table}

\subsubsection{Evaluation Metrics} We adopt \emph{Target Accuracy (TA)}, \emph{Stolen Accuracy (SA)},  and \emph{Number of Queries (\#Queries)} as evaluation metrics. TA (or SA) is the testing accuracy of a downstream classifier trained and tested using a target (or stolen) encoder as feature extractor.  
\#Queries is the number of queries to the EaaS API. Suppose we have a downstream task whose testing dataset is $\mathcal{D}_d=\{(\mathbf{x}_i,y_i)\}_{i=1}^n$. Moreover, we use $g_t$ and $g_s$ to respectively denote the downstream classifier built upon the target encoder $f_t$ and the stolen encoder $f_s$. Formally, we define the metrics as follows:

     \myparatight{Target Accuracy (TA)} TA is  the fraction of testing examples in the testing dataset $\mathcal{D}_d$ that are correctly classified by the downstream classifier $g_t$ when the target encoder is used as a feature extractor. Formally, we have the following: $TA = \frac{\sum_{(\mathbf{x}_i,y_i)\in\mathcal{D}_d}\mathbb{I}(g_t \circ f_t (\mathbf{x}_i)=y_i)}{|\mathcal{D}_d|}$, 
    where $g_t\circ f_t$  is the composition of the target encoder $f_t$ and the downstream classifier $g_t$, and $\mathbb{I}$ is an indicator function.
    
    \myparatight{Stolen Accuracy (SA)} SA is  the fraction of testing examples that are correctly classified by the downstream classifier $g_s$. Formally, we have: $SA = \frac{\sum_{(\mathbf{x}_i,y_i)\in\mathcal{D}_d}\mathbb{I}( g_s \circ f_s (\mathbf{x}_i)=y_i)}{|\mathcal{D}_d|}$, 
    where $g_s \circ f_s$ is the composition of  $f_s$ and  $g_s$. 
   
   \myparatight{Number of Queries (\#Queries)} \#Queries is the number of queries that are sent by an attacker to the EaaS API of a target encoder in order to steal it. Note that \#Queries of {\name} is the size of the surrogate dataset. Therefore, we omit  \#Queries in  most of our experimental results for simplicity.

\begin{figure*}[!t]
	 \centering
\subfloat[STL10]{\includegraphics[width=0.32\textwidth]{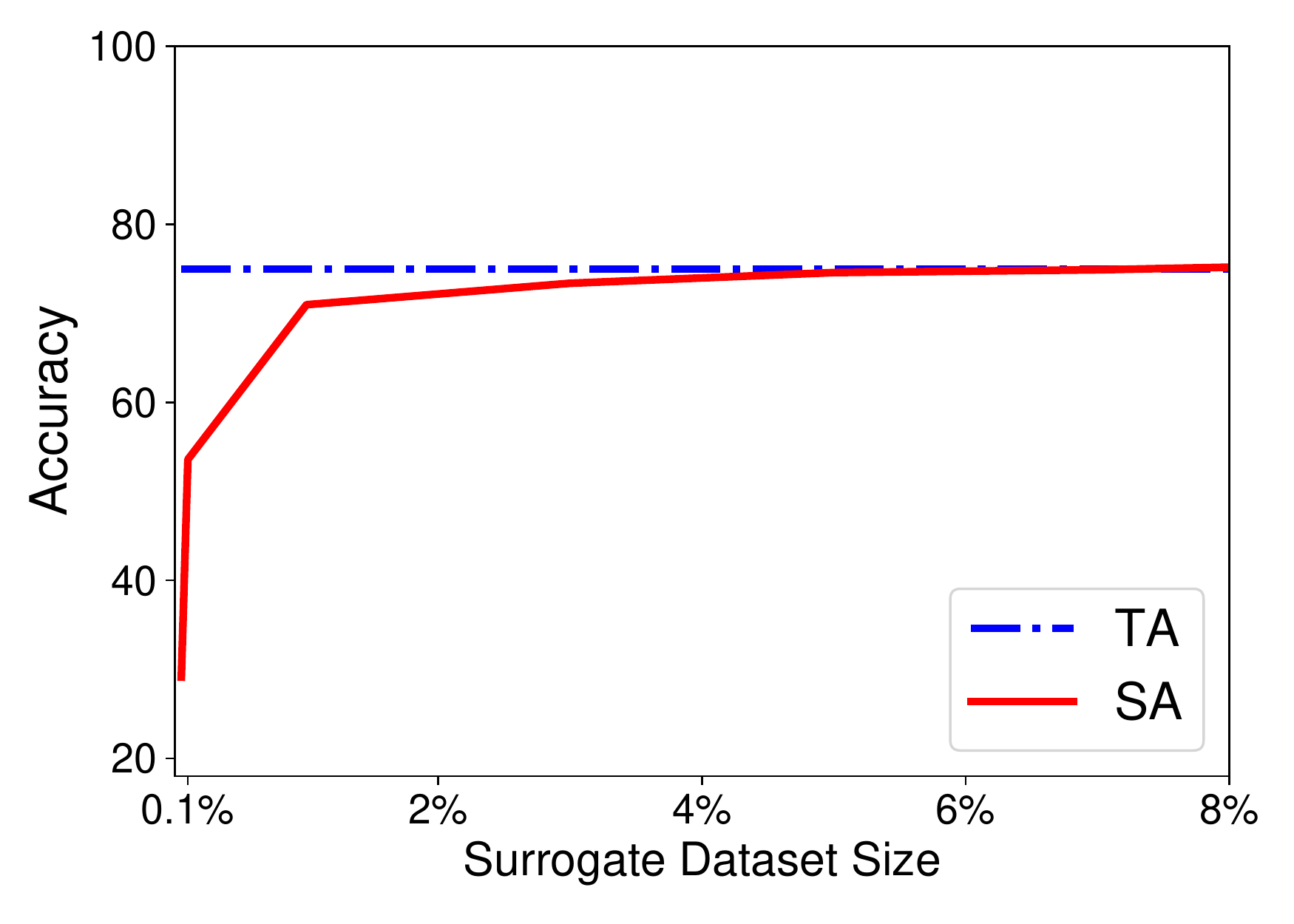}}
\subfloat[Food101]{\includegraphics[width=0.32\textwidth]{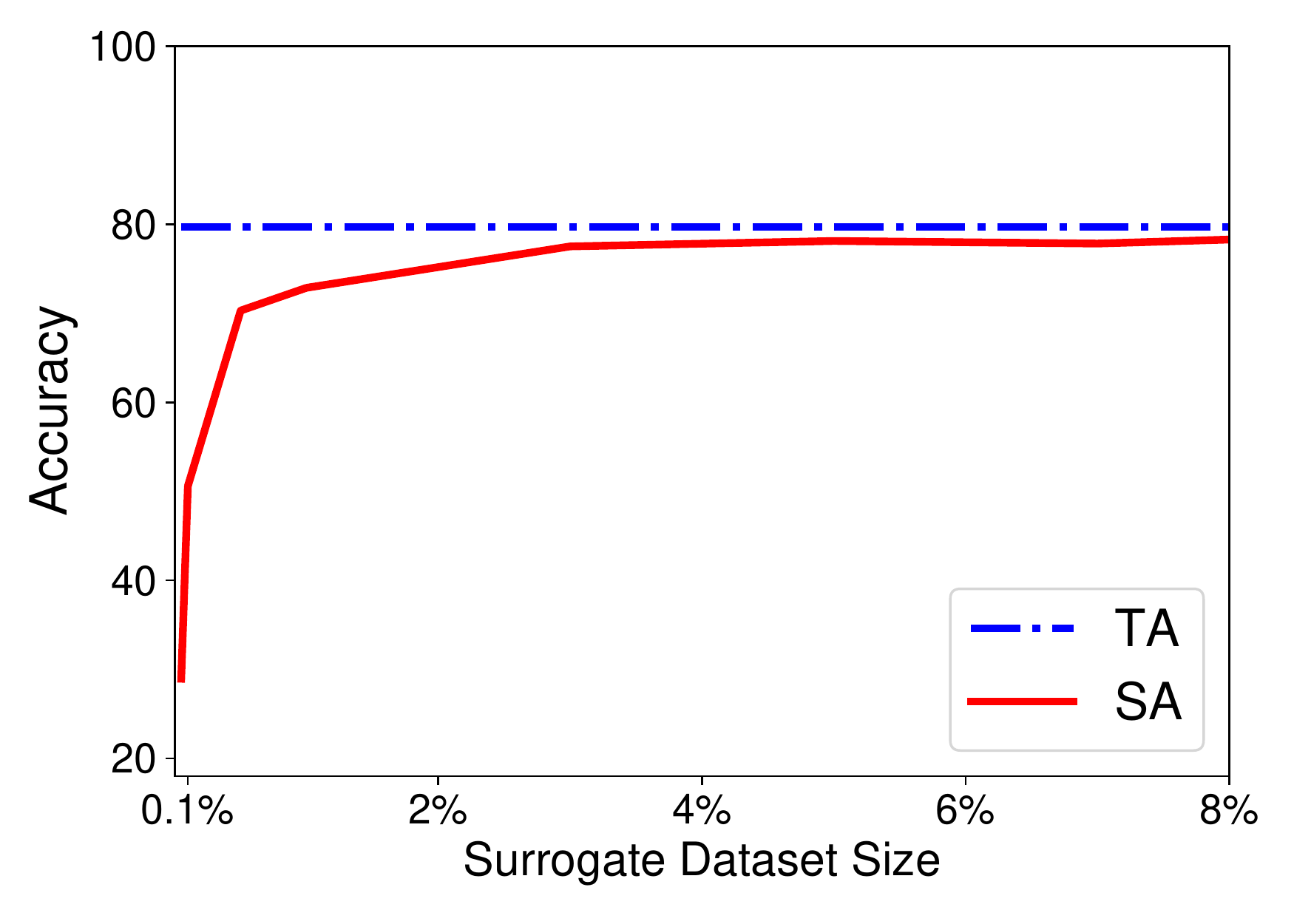}}
\subfloat[CIFAR10]{\includegraphics[width=0.32\textwidth]{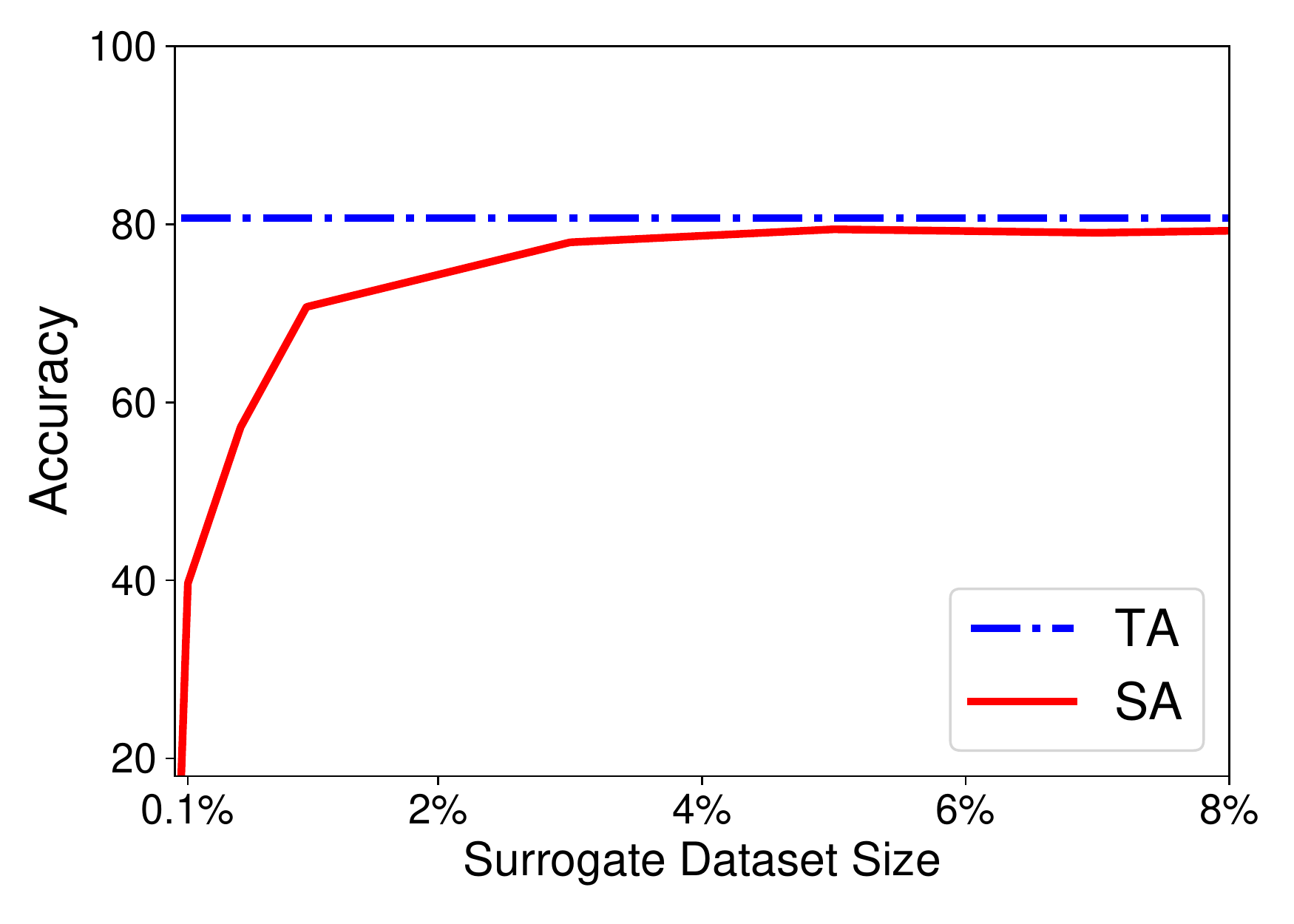}}
\vspace{-2mm}
\caption{Impact of the surrogate dataset size on SA of {\name} for the three pre-training datasets.}
\label{impact_of_query_number}
\end{figure*}

\begin{table}[!t]\renewcommand{\arraystretch}{1.2} 
 	\fontsize{8}{8}\selectfont
	\centering
	\caption{{\name} requires less computation resource than pre-training a target encoder from scratch. }
	\setlength{\tabcolsep}{1mm}
	{
	\begin{tabular}{|c|c|c|}
		\hline
	\makecell{Pre-training \\Dataset} & \makecell{Pre-training\\Time (hrs)} & \makecell{StolenEncoder\\Training Time (hrs)} \\  \hline
	\multirow{1}{*}{STL10} & 26.27 & 0.28 \\ \cline{1-3}  \hline
	\multirow{1}{*}{Food101} & 22.73 & 0.23  \\ \cline{1-3} \hline
	\multirow{1}{*}{CIFAR10} & 12.50 & 0.12\\ \cline{1-3} 

	\end{tabular}
	}
	\label{default_time_table}
	\vspace{-3mm}
\end{table}

\subsection{Experimental Results}

\myparatight{{\name} is effective} Table~\ref{default_table} shows the TA, SA, and the ratio between SA and TA (i.e., SA/TA) for different downstream datasets when the target encoders are pre-trained on STL10, Food101, and CIFAR10. As the results show, {\name} maintains the functionality of the target encoders. 
In particular, the ratios SA/TA are no smaller than 0.98 in all cases and even larger than 1 when the downstream dataset is SVHN. We suspect the reason that the ratios on SVHN are larger than 1 is that SVHN is more similar to the surrogate dataset (i.e., ImageNet dataset) than the three pre-training datasets, and thus the stolen encoders extract more distinguishable features for SVHN than the target encoders. 
Our experimental results demonstrate that {\name} can effectively steal the functionality of a target encoder. 

Table~\ref{default_time_table} shows the times of pre-training the target encoders and training the stolen encoders in {\name}. Our results show that training a stolen encoder in {\name} is two orders of magnitude more computationally efficient than pre-training a target encoder from scratch. Note that the surrogate dataset size is only 5\% of the pre-training dataset size in these experiments. Therefore, training a stolen encoder  requires much less data and computation resources than pre-training a target encoder from scratch.

\myparatight{Impact of $\lambda$} Figure~\ref{impact_of_lambda} shows the impact of $\lambda$ on SA of {\name} for the three pre-training datasets, where the downstream dataset is GTSRB. Note that TA is a constant for each pre-training dataset when we vary $\lambda$.  
We observe that  SA first increases and then saturates as $\lambda$ increases. Our results show that {\name}  achieves good performance with a large range of $\lambda$. In particular, we can set $\lambda$ to be a reasonably large value (e.g., 20) in practice.

\begin{table}[tp]\renewcommand{\arraystretch}{1.2} 
 	\fontsize{8}{8}\selectfont
	\centering
	\caption{Impact of the surrogate dataset distribution on SA of {\name}. } 
	\setlength{\tabcolsep}{1mm}
	{
	\begin{tabular}{|c|c|c|}
		\hline
	\makecell{Pre-training \\Dataset} & \makecell{Surrogate Dataset} & \makecell{SA (\%)}  \\ \hline \hline
	\multirow{3}{*}{STL10}
	& \makecell{a subset of pre-training dataset} & 73.82   \\ \cline{2-3}  
	& \makecell{same dist. as pre-training dataset} & 73.81  \\ \cline{2-3}  
	& \makecell{diff. dist. from pre-training dataset} & 74.57  \\ \cline{1-3} \hline \hline
	\multirow{3}{*}{Food101}
	& \makecell{a subset of pre-training dataset} & 78.20   \\ \cline{2-3}  
	& \makecell{same dist. as pre-training dataset} & 77.95   \\ \cline{2-3}  
	& \makecell{diff. dist. from pre-training dataset} & 78.12   \\ \cline{1-3} \hline \hline
	\multirow{3}{*}{CIFAR10}
	& \makecell{a subset of pre-training dataset} & 78.81   \\ \cline{2-3}  
	& \makecell{same dist. as pre-training dataset} & 78.84   \\ \cline{2-3} 
	& \makecell{diff. dist. from pre-training dataset} & 79.43   \\ \cline{1-3}
	\end{tabular}
	}
	\label{impact_of_surrogate_data}
 	\vspace{-3mm}
\end{table}

\myparatight{Impact of surrogate dataset}  We explore the impact of both the size and distribution of a surrogate dataset on {\name}.  Figure~\ref{impact_of_query_number} shows the impact of the surrogate dataset size on SA of {\name}, where the surrogate dataset is sampled from  ImageNet \CR{uniformly at random without replacement} and its size is calculated as the fraction of the pre-training dataset size.  Note that the size of the surrogate dataset is also the \#Queries incurred by {\name}.  
Naturally, SA first increases and then saturates as the size of the surrogate dataset increases. Moreover, {\name} can already achieve SAs close to TAs when the size of the surrogate dataset is small, e.g.,  3\% of the pre-training dataset size. 

Table~\ref{impact_of_surrogate_data} shows the SAs of {\name} when the surrogate dataset has different distributions.  In particular, we consider three scenarios. In the scenario ``a subset of pre-training dataset'', we randomly select images from the pre-training dataset as the surrogate dataset. In the scenario ``same dist. as pre-training dataset'', we randomly select images from the testing dataset of STL10 (or Food101 or CIFAR10) as the surrogate dataset when the pre-training dataset is STL10 (or Food101 or CIFAR10). In other words, the surrogate dataset does not overlap with the pre-training dataset but they  have the same distribution. In the scenario ``diff. dist. from pre-training dataset'' (our default setting), we randomly select images from the ImageNet dataset as the surrogate dataset. We find that {\name} achieves similar SAs in the three scenarios for the same pre-training dataset, which implies that an attacker can use {\name} to effectively steal a target encoder without the distribution of its pre-training dataset.

\begin{table}[tp]\renewcommand{\arraystretch}{1.3}

  \setlength{\tabcolsep}{+1pt}
  \centering
  \fontsize{7.5}{8}\selectfont
  \caption{Impact of the target encoder architecture, where the stolen encoder architecture is ResNet34. } 
  \begin{tabular}{|c|c|c|c|c|c|c|}
    \hline
    \multirow{3}{*}{\makecell{Target Encoder\\Architecture \\(\#Parameters)}} &
      \multicolumn{6}{c|}{Pre-training Dataset} \cr\cline{2-7}
    &  \multicolumn{2}{c|}{STL10 }& \multicolumn{2}{c|}{Food101} & \multicolumn{2}{c|}{CIFAR10}  \\ \cline{2-7}
    & TA (\%) &SA (\%)&TA (\%)&SA (\%)&TA (\%)&SA (\%) \\ \hline 
    ResNet34 (21.3M) & 76.85 & 75.34 & 79.44 & 78.17 & 76.58 & 76.76  \\ \hline
    VGG19\_bn (20.0M)  & 40.91 & 45.46 & 43.38 &48.80 & 42.68 & 48.52   \\ \hline
    VGG19 (20.0M)    & 21.07 & 26.40 & 24.81 &32.78 & 26.88 &41.69 \\ \hline
    ResNet18 (11.2M) & 74.97 & 74.57 & 79.69 & 78.12 & 80.67 & 79.43  \\ \hline
    DenseNet121 (7.5M)  & 57.21 & 60.75 & 60.46 & 61.78 & 59.78 & 61.69   \\ \hline
    MobileNetV2 (4.0M)  & 38.26 & 46.33 & 40.71 & 44.78 & 38.75 &47.26  \\ \hline
    ShuffleNetV2x1 (1.8M) & 45.10 & 48.43 & 43.36 & 48.61 & 45.60 & 49.79  \\ \hline
  \end{tabular}
  \label{impact_of_arch}
  \vspace{-3mm}
\end{table}

\myparatight{Impact of encoder architecture} An attacker can select an expressive/powerful architecture for the stolen encoder regardless of the target encoder architecture. Table~\ref{impact_of_arch} shows the experimental results when  the stolen encoder architecture is ResNet34 and the target encoder uses different architectures. Our results show that {\name} is effective for different target encoder architectures as the SAs are close to or higher than the corresponding TAs. 
We also found that if an attacker selects an architecture (e.g., MobileNetV2) that is less expressive than the target encoder architecture (e.g., ResNet34), then the SAs are smaller than TAs.

\begin{table}[!t]\renewcommand{\arraystretch}{1.2} 
 	\fontsize{8}{8}\selectfont
	\centering
	\caption{Impact of the distance metric on SA. } 
	\setlength{\tabcolsep}{1mm}
	{
	\begin{tabular}{|c|c|c|}
		\hline
	\makecell{Pre-training Dataset} & \makecell{Distance Metric} & \makecell{SA (\%)}  \\ \hline \hline
	\multirow{3}{*}{STL10}
	& \makecell{cosine distance} & 59.11   \\ \cline{2-3}  
	& \makecell{$\ell_{1}$ distance} & 72.82  \\ \cline{2-3}  
	& \makecell{$\ell_{2}$ distance} & 74.57  \\ \cline{1-3}  \hline \hline
	\multirow{3}{*}{Food101}
	& \makecell{cosine distance} & 61.27   \\ \cline{2-3}  
	& \makecell{$\ell_{1}$ distance} & 76.55   \\ \cline{2-3}  
	& \makecell{$\ell_{2}$ distance} & 78.12   \\ \cline{1-3}  \hline \hline
	\multirow{3}{*}{CIFAR10}
	& \makecell{cosine distance} & 61.07   \\ \cline{2-3}  
	& \makecell{$\ell_{1}$ distance} & 74.52   \\ \cline{2-3} 
	& \makecell{$\ell_{2}$ distance} & 79.43   \\ \cline{1-3}
	\end{tabular}
	}
	\label{impact_of_distance_metric}
 	\vspace{-3mm}
\end{table}

\myparatight{Impact of distance metric} Table~\ref{impact_of_distance_metric} shows  the impact of distance metric on SA of {\name} for different pre-training datasets, where the downstream dataset is GTSRB. Cosine distance measures the minus cosine similarity of the angle between two feature vectors, $\ell_1$ distance is the sum of the absolute difference of each dimension between two feature vectors, while $\ell_2$ distance is the standard Euclidean distance. 
First,  both $\ell_2$ distance and $\ell_1$ distance achieve larger SAs than cosine distance. The reason is that cosine distance only measures the angle between two feature vectors while ignoring their magnitudes. Therefore, it is  likely that the feature vector outputted by a stolen encoder for an input image is substantially different from that outputted by a target encoder even if the cosine distance is -1. Second, $\ell_2$ distance achieves larger SAs than  $\ell_1$ distance. We suspect the reason is that $\ell_2$ distance  makes  the difference of each dimension between the feature vectors outputted by the target encoder and stolen encoder small, while  $\ell_1$ distance aims to make the differences between two feature vectors sparse, i.e., some dimensions have 0 difference but other dimensions may have large differences.

\myparatight{Impact of contrastive learning algorithm used to pre-train a target encoder} Table~\ref{impact_of_cl_algorithm} shows the TAs and SAs when MoCo and SimCLR are used to pre-train target encoders on the three pre-training datasets, where the downstream dataset is GTSRB. Our experimental results show that {\name} is effective for both MoCo and SimCLR.

\myparatight{Impact of data augmentation operations} {\name} uses augmented images from the surrogate dataset when training a stolen encoder. We consider the 16 combinations of four data augmentation operations including  RandomResizedCrop,  RandomHorizontalFlip,  ColorJitter, and RandomGrayScale, which are used to pre-train a target encoder. Table~\ref{impact_of_aug} in Appendix shows the SAs of {\name} for the 16 combinations, where the pre-training dataset is CIFAR10 and downstream dataset is GTSRB. We find that {\name} with any combination of the four data augmentation operations achieves high SAs, which indicates that an attacker does not need to know the data augmentation operations used to pre-train the target encoder.

\myparatight{Impact of other parameters} We study the impact of the parameters (i.e., \#epochs and learning rate) in pre-training a target encoder, the parameters (i.e., \#epochs, learning rate, and batch size) in training a stolen encoder, and the parameters (\#epochs, learning rate, and \#neurons in hidden layers) in training a downstream classifier. When studying the impact of one parameter, we fix other parameters to the default settings. Table~\ref{result_of_different_parameters} in Appendix shows the results when the pre-training dataset is CIFAR10 and downstream dataset is GTSRB. Our results show that {\name} is effective for a wide range of parameter settings.

\begin{table}[tp]\renewcommand{\arraystretch}{1.2} 
 	\fontsize{8}{8}\selectfont
	\centering
	\caption{Impact of the contrastive learning algorithm used to pre-train a target encoder on {\name}. } 
	\setlength{\tabcolsep}{1mm}
	{
	\begin{tabular}{|c|c|c|c|}
		\hline
	\makecell{Pre-training \\Dataset} & \makecell{Target Encoder\\ Algorithm} & \makecell{TA (\%)} & \makecell{SA (\%)}  \\ \hline \hline
	\multirow{2}{*}{STL10}
	& \makecell{MoCo} & 80.36 & 83.50   \\ \cline{2-4}  
	& \makecell{SimCLR} & 74.97 & 74.57  \\ \cline{1-4} \hline \hline
	\multirow{2}{*}{Food101}
	& \makecell{MoCo} & 81.97 & 80.91   \\ \cline{2-4} 
	& \makecell{SimCLR} & 79.69 & 78.12   \\ \cline{1-4} \hline \hline 
	\multirow{2}{*}{CIFAR10}
	& \makecell{MoCo} & 78.92 & 78.41   \\ \cline{2-4}  
	& \makecell{SimCLR} & 80.67 & 79.43   \\ \cline{1-4}
	\end{tabular}
	}
	\label{impact_of_cl_algorithm}
 	\vspace{-4mm}
\end{table}

\myparatight{Comparing {\name} with its variants} Table~\ref{comparision_table} shows comparison results when the surrogate dataset size is 3\% and 5\% of the pre-training dataset size,  where the pre-training dataset is CIFAR10 and the downstream dataset is GTSRB.  Our results show that {\name} outperforms Pre-training-encoder, which means that an attacker can obtain a better encoder by using {\name} than by pre-training on its surrogate dataset locally.  {\name} outperforms StolenEncoder-w/o-aug, which implies that  augmenting the surrogate dataset by data augmentations improve the functionality of the stolen encoder. Moreover, {\name} achieves comparable functionality with StolenEncoder-query-aug but requires much less queries, which means that approximating the feature vector of an augmented image as that of the original image produced by the target encoder (i.e., approximating $\mathcal{L}_2'$ as $\mathcal{L}_2$) does not sacrifice functionality of the stolen encoder but reduces query cost substantially.

\begin{table}[tp]\renewcommand{\arraystretch}{1.2} 
% \vspace{2mm}
	\fontsize{8}{8}\selectfont
	\centering
	\caption{Comparing variants of {\name}.  The pre-training dataset is CIFAR10 and downstream dataset is GTSRB.  }
	\setlength{\tabcolsep}{1mm}
	{
	\begin{tabular}{|c|c|c|c|}
		\hline
	\makecell{Surrogate \\Dataset Size} & \makecell{Method} & \makecell{SA (\%)} & \makecell{\#Queries}  \\ \hline \hline
	\multirow{4}{*}{3\%}
	& Pre-training-encoder & 66.42  & 0  \\ \cline{2-4}  
	& StolenEncoder-w/o-aug & 44.47  & 1,500  \\ \cline{2-4} 
	& StolenEncoder-query-aug & 76.71  & 151,500  \\ \cline{2-4}  
	& StolenEncoder & 77.73  & 1,500  \\ \cline{1-4}   \hline \hline
	\multirow{4}{*}{5\%}
	& Pre-training-encoder & 71.40 & 0  \\ \cline{2-4}  
	& StolenEncoder-w/o-aug & 51.50  & 2,500  \\ \cline{2-4} 
	& StolenEncoder-query-aug & 79.11  & 252,500  \\ \cline{2-4}  
	& StolenEncoder & 79.17  & 2,500  \\ \cline{1-4} 
	\end{tabular}
	}
	\label{comparision_table}
 	\vspace{-2mm}
\end{table}

\section{Case Studies}
\label{sec:casestudy}

We evaluate our {\name} on three real-world pre-trained image encoders. Particularly, one is  pre-trained on the ImageNet dataset by Google~\cite{chen2020simple},  one  is the  CLIP image encoder pre-trained by OpenAI~\cite{radford2021learning}, and one is the Clarifai's General Embedding encoder~\cite{clarifai_image_embedding}. The ImageNet and CLIP encoders are publicly available, while the Clarifai encoder is deployed as a paid EaaS.  
 We acknowledge that  it is not practically relevant to steal the ImageNet and CLIP image encoders, as they are publicly available. However, tech companies may set their future real-world image encoders to be private and only expose EaaS APIs for customers due to various reasons such as intellectual property protection and ethical concerns. An example is that OpenAI publicly released its pre-trained GPT and GPT-2 models, but only provides EaaS API for its pre-trained GPT-3 model due to economic and ethical considerations (e.g., GPT-3 may be misused to synthesize fake news)~\cite{OpenAIAPI}. Although these GPT models are text encoders, a similar trend may happen for image encoders, especially when they become really powerful and have a wide range of real-world (good or bad) applications.  Therefore, we perform experiments on the ImageNet and CLIP image encoders to show the effectiveness of  {\name} on real-world, large-scale image encoders.

\subsection{Experimental Setup}
%\vspace{-2mm}
\myparatight{Pre-training datasets and target encoders} Both the ImageNet and CLIP encoders are pre-trained by contrastive learning. Specifically, the ImageNet target encoder was pre-trained using SimCLR on the ImageNet dataset, which has 1.3 million images. The CLIP target encoder is pre-trained on a dataset with 400 million image-text pairs collected from the Internet. The architectures of the two target encoders are  ResNet50 with 2,048 output dimensions and modified ResNet50 with 1,024 output dimensions, respectively. The input size of both target encoders is 224 $\times$ 224 $\times$ 3. During pre-training, the ImageNet target encoder uses data augmentation operations including RandomResizedCrop, ColorJitter, RandomGrayScale, and RandomHorizontalFlip, while the CLIP target encoder only uses  RandomResizedCrop. The Clarifai encoder is deployed as a paid EaaS, which charges \$3.2 per 1,000 queries~\cite{ClarifaiPriceSheet}. The technique used to pre-train the Clarifai encoder, its input size, its pre-training dataset, and its architecture are not publicly known.

We note that stealing the Clarifai encoder does not influence other users of the Clarifai EaaS. Prior work~\cite{tramer2016stealing,orekondy2019knockoff} on classifier stealing attacks also performed evaluation on real-world classifier APIs. Moreover, we have notified the Clarifai provider about its vulnerability to stealing attacks.      

\myparatight{Attack settings} 
In our experiments in Section~\ref{sec:exp}, we sample images from  ImageNet as the surrogate dataset. Since the ImageNet target encoder is pre-trained on ImageNet, we construct surrogate dataset from a different data source. In particular, we construct surrogate dataset  from  STL10, which includes 113,000 images in total. We resize each image in STL10 to 224 $\times$ 224 $\times$ 3 to fit the input size of the target encoders. In the ImageNet target encoder experiments, we assume the surrogate dataset size is 5\% of the pre-training dataset size, i.e.,  the surrogate dataset includes 64,058 images sampled from STL10 uniformly at random.   In the CLIP target encoder experiments, we use the entire STL10 dataset as the surrogate dataset. We note that the surrogate dataset size in the CLIP experiments is less than 0.03\% (113K/400M) of the pre-training dataset size. 
In the Clarifai experiments, we sample 5,000 images from STL10 uniformly at random as the surrogate dataset. 
We use the default settings in Section~\ref{sec:exp} for the other parameters of {\name}. 

\myparatight{Downstream datasets and downstream classifiers} For the downstream datasets and downstream classifiers, we use the default settings in Section~\ref{sec:exp} except that  we resize all images to 224 $\times$ 224 to fit the input dimension of the  ImageNet and CLIP target encoders. The Clarifai automatically resizes each image to its input size. 

\begin{table}[tp]\renewcommand{\arraystretch}{1.2} 
 	\fontsize{8}{8}\selectfont
	\centering
	\caption{{\name} is effective for the real-world ImageNet and CLIP target encoders as well as the Clarifai's General Embedding encoder. }
	\setlength{\tabcolsep}{1mm}
	{
	\begin{tabular}{|c|c|c|c|c|}
		\hline
	\makecell{Target \\Encoder} & \makecell{Downstream \\Dataset} & \makecell{TA (\%)} & \makecell{SA (\%)} &  $\frac{\text{SA}}{\text{TA}}\times 100\%$ \\ \hline \hline
	\multirow{4}{*}{ImageNet}
	& MNIST & 98.18 & 98.70 & 101  \\ \cline{2-5}  
	& FashionMNIST & 91.49 & 89.91 & 98  \\ \cline{2-5}  
	& SVHN & 72.73 & 78.56 & 108  \\ \cline{2-5}  
	& GTSRB & 75.29  & 76.96 & 102  \\ \hline \hline
	\multirow{4}{*}{CLIP}
	& MNIST & 98.66 & 96.56 & 98  \\ \cline{2-5}  
	& FashionMNIST & 90.37 & 84.75 & 94  \\ \cline{2-5}  
	& SVHN & 70.87 & 71.27 & 101  \\ \cline{2-5}   
	& GTSRB & 80.46 & 74.50 & 93  \\ \hline \hline  
	\multirow{4}{*}{\makecell{Clarifai}}
	& MNIST & 97.39 & 97.04 & 100  \\ \cline{2-5}  
	& FashionMNIST & 89.57 & 88.15 & 98  \\ \cline{2-5}  
	& SVHN & 67.21 & 68.42 & 102  \\ \cline{2-5}   
	& GTSRB & 61.00 & 58.16 & 95  \\ \cline{1-5}  
	\end{tabular}
	}
	\label{casestudies}
	 \vspace{-4mm}
\end{table}

\subsection{Experimental Results}
%\vspace{-2mm}
\myparatight{TAs and SAs} Table~\ref{casestudies} shows the TA,  SA, and the ratio SA/TA when stealing the three real-world target encoders.  
Our results show that {\name} is  effective for real-world target encoders as the ratios SA/TA are at least 93\%.   

\myparatight{Economic cost} In the Clarifai experiments, training and testing the downstream classifiers using the paid EaaS API costs us \$931.6. However, stealing the Clarifai encoder using 5,000 queries only costs us \$16. Our results have two implications. First, an attacker can steal the encoder, and then train and test its downstream classifiers using the stolen encoder, which substantially saves economic cost for the attacker with no or minor degradation of the downstream classifiers' accuracy. Second, an attacker can economically benefit from stealing the encoder and deploying it as its own EaaS, which customers can query to train/test downstream classifiers.

\myparatight{Data and computation resources} For the ImageNet target encoder, the surrogate dataset size is 5\% of the pre-training dataset size; for the CLIP target encoder, the surrogate dataset size is less than 0.03\% of the pre-training dataset size; and for the Clarifai target encoder, the surrogate dataset size is 5,000 (the pre-training dataset size is not publicly known). Moreover, Table~\ref{case_study_time_table} shows the hardware and times used to pre-train the target encoders and train the stolen encoders in {\name}, where the pre-training hardware and times of the ImagetNet and CLIP target encoders are obtained from the corresponding papers, while the pre-training hardware and time of the Clarifai target encoder are not publicly known.  Our results show that stealing a real-world target encoder using {\name} requires much less data and computation resources than pre-training the target encoder from scratch. In other words, a less resourceful attacker can use {\name} to steal an encoder that is pre-trained by a much more resourceful EaaS provider.

\begin{table}[!t]\renewcommand{\arraystretch}{1.2}

\addtolength{\tabcolsep}{-2pt}
  \centering
  \fontsize{8}{8}\selectfont
  \caption{{\name} requires much less computation resources than pre-training the real-world target encoders from scratch.} 
  \begin{tabular}{|c|c|c|c|c|}
    \hline
    \multirow{2}{*}{\makecell{Target \\Encoder}} 
    &  \multicolumn{2}{c|}{Pre-training }& \multicolumn{2}{c|}{StolenEncoder}  \\ \cline{2-5}
    & Hardware & Time (hrs) & Hardware &Time (hrs) \\ \hline 
    ImageNet & \makecell{ TPU v3} & 192  & \multirow{3}{*}{\makecell{Quadro\\RTX-6000 GPU}}  & 30.4 \\ \cline{1-3}\cline{5-5}
    CLIP  & \makecell{V100 GPU}  & 255,744  &   & 53.9  \\ \cline{1-3}\cline{5-5}
    Clarifai  & - & - &  & 0.1 \\ \hline
  \end{tabular}
  \label{case_study_time_table}
 \vspace{-4mm}
\end{table}

\section{Defenses}
We generalize defenses against classifier stealing attacks to defend against {\name}. In particular, one popular category of defenses aim to perturb a confidence score vector predicted by the target classifier for a query before returning it to a customer~\cite{tramer2016stealing,orekondy2019knockoff,orekondy2020prediction} (for a comprehensive discussion on defenses against classifier stealing attacks, please refer to Section~\ref{relatedwork}). The intuition is that an attacker can only train a less accurate stolen classifier using the perturbed confidence score vectors. 

We generalize three such defenses to defend against {\name}. In \emph{top-$k$ features}, the EaaS API resets the features, whose absolute values are not the top-$k$ largest, to 0 before returning a feature vector to a customer; in \emph{feature rounding}, the EaaS API returns rounded features to a customer; and in \emph{feature poisoning}, the EaaS API adds carefully crafted perturbation to a feature vector to poison the training of a stolen encoder.  
In all these defenses,  an attacker uses the perturbed feature vectors to train a stolen encoder, i.e., the attacker solves the optimization problem in Equation~(\ref{final_loss}) using the perturbed feature vectors returned by the EaaS API. Moreover, the downstream classifiers are also trained and tested using the perturbed feature vectors.

\subsection{Top-$k$ Features} In this defense, given a query, the EaaS API first uses the target encoder to calculate its feature vector; then the API resets the features, whose absolute values are not the top-$k$ largest ones, to be 0; and finally, the API returns the perturbed feature vector. 
Table~\ref{tab:defense_top_k} shows the defense results of top-$k$ features with different values of $k$ for the three pre-training datasets, where the downstream dataset is GTSRB.   Our results show that top-$k$ features are insufficient. In particular,  SAs are close to or higher than TAs no matter what $k$ is used. In other words, 
although a smaller $k$ reduces  SAs,  TAs are also reduced by similar or larger magnitudes. We note that top-$k$ features are generalized from top-$k$ confidence scores, which was first explored to defend against membership inference attacks~\cite{shokri2017membership,jia2019memguard} and then extended to classifier stealing attacks~\cite{orekondy2019knockoff}. 

\begin{table}[tp]\renewcommand{\arraystretch}{1.2}

% \addtolength{\tabcolsep}{-2pt}
  \centering
  \fontsize{8}{8}\selectfont
  \caption{Defense results of top-$k$ features. } 
  \begin{tabular}{|c|c|c|c|c|c|c|}
    \hline
    \multirow{3}{*}{\makecell{$k$}} &
      \multicolumn{6}{c|}{Pre-training Dataset} \cr\cline{2-7}
    &  \multicolumn{2}{c|}{STL10 }& \multicolumn{2}{c|}{Food101} & \multicolumn{2}{c|}{CIFAR10}  \\ \cline{2-7}
    & TA (\%) &SA (\%)&TA (\%)&SA (\%)&TA (\%)&SA (\%) \\ \hline 
    512 & 74.97 & 74.57 & 79.69 & 78.12 & 80.67 & 79.43 \\ \hline
    200  & 74.97 & 75.10 & 79.32 & 78.27 & 80.62 & 78.86   \\ \hline
    100  & 72.85 & 74.87 & 76.35 & 78.29 & 78.07 & 78.66 \\ \hline
    50 & 67.93 & 72.68 & 69.35 & 77.14 & 72.57 & 74.48 \\ \hline
    30 & 60.99 & 69.77 & 62.91 & 75.84 & 68.24 & 69.49   \\ \hline
    10 & 48.67 & 63.44 & 47.40 & 61.30 & 56.59 & 61.39  \\ \hline
    5 & 41.06 & 47.76 & 36.36 & 28.16 & 47.77 & 51.59  \\ \hline
    1 & 18.59 & 17.58 & 16.65 & 17.58 & 26.28 & 21.06 \\ \hline
  \end{tabular}
  \label{tab:defense_top_k}
%   \vspace{5mm}
\end{table}

\begin{table}[tp]\renewcommand{\arraystretch}{1.2}

% \addtolength{\tabcolsep}{-2pt}
  \centering
  \fontsize{8}{8}\selectfont
  \caption{Defense results of feature rounding.} 
  \begin{tabular}{|c|c|c|c|c|c|c|}
    \hline
    \multirow{3}{*}{\makecell{$m$}} &
      \multicolumn{6}{c|}{Pre-training Dataset} \cr\cline{2-7}
    &  \multicolumn{2}{c|}{STL10 }& \multicolumn{2}{c|}{Food101} & \multicolumn{2}{c|}{CIFAR10}  \\ \cline{2-7}
    & TA (\%) &SA (\%)&TA (\%)&SA (\%)&TA (\%)&SA (\%) \\ \hline 
    1 & 66.25 & 75.10 & 70.13 & 78.73 & 72.87 & 77.05 \\ \hline
    2  & 74.62 & 75.69 & 79.28 & 78.08 & 80.50 & 79.08   \\ \hline
    3  & 75.23 & 76.47 & 79.43 & 78.83 & 80.82 & 78.59 \\ \hline
  \end{tabular}
  \label{tab:defense_keep_m}
%  \vspace{5mm}
\end{table}

\subsection{Feature Rounding} In this defense, the EaaS API  rounds each feature to $m$ decimals. For instance, rounding a feature value 0.0123 to 2 decimals results in a feature value 0.01. The intuition is that, with coarser-grained feature vectors, an attacker may be able to train an inferior stolen encoder. Table~\ref{tab:defense_keep_m} shows the defense results of feature rounding with different values of $m$ for the three pre-training datasets. Our results show that  
feature rounding is ineffective at defending against {\name}. Specifically, SAs of {\name} are close to TAs even if the EaaS API rounds each feature to 1 decimal. Moreover, we find that both TAs and SAs are not sensitive to the granularity of the feature values, as they are similar when $m=1,2,3$.  We note that feature rounding is generalized from  confidence score rounding, which was explored to defend against model inversion attacks~\cite{fredrikson2015model}, classifier stealing attacks~\cite{tramer2016stealing}, and hyperparameter stealing attacks~\cite{wang2018stealing}. 

\subsection{Feature Poisoning} 

This defense is generalized from confidence score poisoning, which was explored to defend against membership inference attacks~\cite{jia2019memguard} and classifier stealing attacks~\cite{orekondy2020prediction}. In particular, the key idea is to add carefully crafted perturbation to a feature vector such that the perturbed feature vectors become a ``data poisoning attack'' to the training of a stolen encoder, i.e., the stolen encoder has inferior functionality. Recall that {\name} minimizes the loss in Equation~(\ref{final_loss}) to train a stolen encoder. In feature poisoning, the API adds a perturbation to the feature vector of any query to maximize the loss in Equation~(\ref{final_loss}). Formally, the API finds a perturbation $\delta$ for each query $\mathbf{x}$ via solving the following optimization problem:  
% {\footnotesize
\begin{align}
\label{perturbation_loss_term_1}
\max_{\delta} & \ d(f_t(\mathbf{x}) + \delta, f_s(\mathbf{x})) + \lambda \cdot d(f_t(\mathbf{x})+\delta,  f_s(\mathcal{A}(\mathbf{x}))), \nonumber\\
s.t. & ||\delta||_p \leq \epsilon,
\end{align}
% }
where $f_t$ is the target encoder, $f_s$ is the attacker's stolen encoder, $d$ is a distance metric, $\lambda$ is a hyperparameter, $\mathcal{A}(\mathbf{x})$ is an augmented version of $\mathbf{x}$, and  $||\delta||_p \leq \epsilon$ means that the $\ell_p$ norm of the perturbation is bounded by $\epsilon$.  $\epsilon$ controls the functionality of the stolen and target encoders. Specifically, a larger $\epsilon$ is expected to degrade the functionality of both a stolen encoder and the target encoder.

\begin{table}[!t]\renewcommand{\arraystretch}{1.2}

% \addtolength{\tabcolsep}{-2pt}
  \centering
  \fontsize{8}{8}\selectfont
  \caption{Defense results of feature poisoning.} % on the two datasets.}
  \subfloat[$||\delta||_2\leq \epsilon$]{\begin{tabular}{|c|c|c|c|c|c|c|}
    \hline
    \multirow{3}{*}{\makecell{$\epsilon$}} &
      \multicolumn{6}{c|}{Pre-training Dataset} \cr\cline{2-7}
    &  \multicolumn{2}{c|}{STL10}& \multicolumn{2}{c|}{Food101} & \multicolumn{2}{c|}{CIFAR10}  \\ \cline{2-7}
    & TA (\%) &SA (\%)&TA (\%)&SA (\%)&TA (\%)&SA (\%) \\ \hline
    0 & 74.97 & 74.57 & 79.69 & 78.12 & 80.67 & 79.43 \\ \hline
    1 & 73.37 & 75.19 & 78.13 & 75.07 & 80.73 & 74.14   \\ \hline
    3 & 73.59 & 73.70 & 74.11 & 72.79 & 82.43 & 71.12  \\ \hline
    5 & 67.51 & 71.19 & 73.49 & 71.70 & 80.44 & 72.85    \\ \hline
    7 & 60.84 & 60.83 & 63.32 & 61.50 & 75.81 & 68.38  \\ \hline
    10& 48.28 & 53.41 & 48.14 & 51.68 & 66.91 & 57.56    \\ \hline
  \end{tabular}
  \label{poisoning_defense_l2_1}}
  
    \subfloat[$||\delta||_\infty\leq \epsilon$ ]{\begin{tabular}{|c|c|c|c|c|c|c|}
    \hline
    \multirow{3}{*}{\makecell{$\epsilon$}} &
      \multicolumn{6}{c|}{Pre-training Dataset} \cr\cline{2-7}
    &  \multicolumn{2}{c|}{STL10}& \multicolumn{2}{c|}{Food101} & \multicolumn{2}{c|}{CIFAR10}  \\ \cline{2-7}
    & TA (\%) &SA (\%)&TA (\%)&SA (\%)&TA (\%)&SA (\%) \\ \hline
    0    & 74.97 & 74.57 & 79.69 & 78.12 & 80.67 & 79.43 \\ \hline
    0.01 & 71.01 & 74.67 & 76.87 & 75.52 & 79.29 & 77.96   \\ \hline
    0.03 & 64.60 & 65.59 & 69.59 & 69.93 & 74.18 & 71.25   \\ \hline
    0.05 & 60.26 & 56.73 & 64.43 & 58.24 & 70.75 & 56.52    \\ \hline
    0.07 & 55.45 & 48.30 & 58.04 & 50.24 & 65.61 & 50.13 \\ \hline
    0.1  & 49.60 & 42.65 & 49.49 & 42.07 & 56.92 & 40.27    \\ \hline
  \end{tabular}
  \label{tab_defense_l_infty}}
\label{featurepoisoning}
 % \vspace{3mm}
\end{table}

It is challenging for the EaaS provider to solve Equation~(\ref{perturbation_loss_term_1}) because it does not have access to the attacker's stolen encoder $f_s$ and the parameter settings used to train it. To give advantages to the service provider, we assume a strong defender, who can use the exactly same parameter settings as the attacker to train a stolen encoder. In particular, the EaaS provider trains a stolen encoder $f_s'$ based on Equation~(\ref{final_loss}) by treating its own pre-training dataset as the surrogate dataset and using the same settings for other parameters as the attacker. Moreover, the EaaS provider replaces the stolen encoder $f_s$ as $f_s'$ in Equation~(\ref{perturbation_loss_term_1}) and solves it for each query to obtain perturbed feature vector, which is returned to a customer/attacker. We assume the parameters including $d$, $\lambda$, and $\mathcal{A}$ in Equation~(\ref{perturbation_loss_term_1}) are the same as those used by the attacker.

 Table~\ref{featurepoisoning}  shows the defense results of feature poisoning with different $\epsilon$'s as well as both $\ell_2$ and $\ell_\infty$ norms to measure the perturbation magnitude. Our results show that feature poisoning is also insufficient to mitigate {\name}. In particular,  although a larger $\epsilon$ decreases the SAs of {\name}, it also decreases the TAs, i.e., feature poisoning degrades the functionality of the stolen encoder by degrading the functionality of the target encoder.

\section{Related Work}
\label{relatedwork}

%\vspace{-2mm}
\myparatight{Stealing classifiers} Most model stealing attacks focused on classifiers~\cite{orekondy2019knockoff,yu2020cloudleak,chandrasekaran2020exploring,tramer2016stealing, jagielski2020high,carlini2020cryptanalytic,kariyappa2021maze, zhu2021hermes}. Roughly speaking, these methods steal the exact model parameters or functionality of target classifiers via querying them~\cite{orekondy2019knockoff,chandrasekaran2020exploring,tramer2016stealing, jagielski2020high,carlini2020cryptanalytic,kariyappa2021maze} or monitoring hardware side-channel information~\cite{zhu2021hermes}. For instance, Tramer et al.~\cite{tramer2016stealing} proposed classifier stealing attacks to logistic regressions, decision trees, support vector machines, and simple neural networks via querying the target classifiers deployed as MLaaS. Orekondy et al.~\cite{orekondy2019knockoff} proposed a reinforcement learning based approach to reduce the number of queries to steal classifiers deployed as MLaaS. Chandrasekaran et al.~\cite{chandrasekaran2020exploring} formally explored the connections between active learning and classifier stealing attacks. Zhu et al.~\cite{zhu2021hermes} proposed a hardware side-channel based attack to steal classifiers. The threat model assumes that the attacker process is running on the same machine as the target classifier, and thus the attacker can monitor the PCI bus. 
 StolenEncoder is different from these work as we aim to steal pre-trained encoders instead of classifiers. 

\CR{A concurrent work~\cite{cong2022sslguard} proposed an attack to steal the parameters of an encoder. In particular, the idea is to make a stolen encoder and a target encoder produce similar feature vectors for inputs in an attacker's surrogate dataset. Their attack is a special case of our attack with $\lambda=0$, i.e., the attacker steals an encoder via minimizing the loss term $\mathcal{L}_1$. As shown in our experiment results in Table~\ref{comparision_table}, our {\name} outperforms such attack. Moreover, they use cosine similarity as the distance metric $d$, which is suboptimal as shown by our experiment results in Table~\ref{impact_of_distance_metric}. We note that knowledge distillation~\cite{hinton2015distilling} can also be generalized to steal a target encoder. In particular, we can treat target encoder + softmax as a teacher classifier and stolen encoder + softmax as a student encoder. Then, we use knowledge distillation to train the student classifier. We found that SAs for StolenEncoder and knowledge distillation (temperature parameter is 1) are respectively 0.7943 and 0.0954 in our default experimental settings. Knowledge distillation does not work well because it uses cross-entropy loss which is designed for classifiers instead of encoders.  }

A few works~\cite{krishna2019thieves,zanella2021grey} proposed attacks to steal  language  models fine-tuned based on a publicly available pre-trained language encoder. In these works, the threat model is that both the target model and the stolen model are fine-tuned based on the same public pre-trained language encoder. For instance, Krishna et al.~\cite{krishna2019thieves} proposed attacks to steal language models fine-tuned based on 
the pre-trained BERT encoder~\cite{devlin2018bert}. In particular, they showed that the attacker can effectively steal the target language model via querying it even if the attacker does not know any training data of the target language model.  Zanella et al.~\cite{zanella2021grey} further improved such stealing attacks via introducing algebraic techniques under grey-box access. These works are different from {\name} because {\name} aims to steal the confidential pre-trained encoder while these works aim to steal a model fine-tuned based on a public pre-trained encoder.

\myparatight{Stealing hyperparameters and training data} Wang and Gong~\cite{wang2018stealing} proposed attacks to steal hyperparameters used to train a model via a black-box access to it, which can be applied to various machine learning algorithms.  Oh et al.~\cite{oh2018towards} proposed attacks to steal hyperparameters of the architecture and optimizer used to train a neural network classifier via querying it. Yan et al.~\cite{yan2020cache} proposed hardware side-channel based attack to steal the architecture of a neural network.  
Their threat model assumes the attacker has access to the same machine where the target classifier is running, and the attacker can monitor the side effects of the CPUs' cache behaviors. He et al.~\cite{he2021stealing} proposed attacks to steal a training graph via querying a graph neural network model that is trained on the graph. They showed that the predictions of a graph neural network classifier leak lots of structural information of the graph used to train the classifier. Jia et al.~\cite{jia2021robust} proposed stealing attacks that assume an untrusted machine learning library is used to train a model. In particular, the untrusted library embeds training data or hyperparameters (e.g., neural network architecture) into the model during training. After the model is deployed, an attacker can query it to steal the embedded training data or hyperparameters.

\myparatight{Defenses against classifier stealing attacks}
Two popular categories of defenses against classifier stealing attacks include 1) detecting malicious/abnormal queries~\cite{juuti2019prada,zhang2021seat}, and 2) perturbing the confidence score vectors predicted by a target classifier for queries~\cite{tramer2016stealing,orekondy2019knockoff,orekondy2020prediction,kariyappa2020defending}. The intuition of the first category of defenses is that the queries (e.g.,  inputs not following the normal testing data distribution, adversarial examples) from an attack may statistically deviate from the normal queries whose distribution is known for the given target classifier, while the intuition of the second category of defenses is that an attacker can only train a less accurate stolen classifier using the perturbed confidence score vectors. It is challenging to apply the first category of defenses to defend against encoder stealing attacks due to two reasons. First, unlike a target classifier, a target encoder is supposed to accept queries from different downstream tasks (i.e., different data distributions), and thus it is intrinsically hard to define ``abnormal'' queries. Second, the attacker can register multiple accounts and distribute its queries among them, making it hard for the service provider to associate and statistically analyze the attacker's queries. In fact, the second reason can also make these defenses ineffective to defend against classifier stealing attacks. In our work, we extend three defenses in the second category to  mitigate {\name}. However, our  results show the insufficiency of these defenses.

We note that watermarking~\cite{adi2018turning,zhang2018protecting,jia2021entangled,lukas2021sok} and fingerprinting~\cite{cao2021ipguard,lukas2021deep} were also proposed to protect the intellectual property of machine learning models. However, these methods aim to verify model ownership after a model has already been stolen.

\myparatight{Other security and privacy issues of pre-trained encoders} Other than stealing attacks, pre-trained encoders are also vulnerable to \CR{data poisoning attacks~\cite{liu2022poisonedencoder}}, backdoor attacks~\cite{jia2021badencoder,carlini2021poisoning}, membership inference attacks~\cite{carlini2021extracting,liu2021encodermi,he2021quantifying}, and possibly other security/privacy attacks~\cite{bommasani2021opportunities,jia202110}.

\section{Conclusion and Future Work}
In this work, we show that an attacker can steal the functionality of a pre-trained image encoder via querying it.  
Such encoder stealing attack can be formulated as a minimization optimization problem, the solution of which is a stolen encoder. Moreover, our extensive evaluation results show that stealing a pre-trained image encoder requires much less data and computation resources than pre-training it from scratch. We also show that defenses generalized from defending against classifier stealing attacks are insufficient to mitigate encoder stealing attacks. Interesting future work includes 1) developing attacks to steal the hyperparameters (e.g., architecture) of a target encoder, and 2) developing new defense mechanisms to mitigate {\name}. 

\myparatight{Acknowledgements} We thank the anonymous reviewers for constructive comments. This work was supported by NSF under Grant No. 1937786 and 2112562.

\begin{table}[!t]\renewcommand{\arraystretch}{1.2}

  \setlength{\tabcolsep}{+6pt}
  \centering

  \caption{Impact of the data augmentation operations used to create augmented images  when training the stolen encoder, where the pre-training dataset is CIFAR10 and downstream dataset is GTSRB. 1: RandomResizedCrop, 2: RandomHorizontalFlip, 3: ColorJitter, and 4: RandomGrayScale. }  
  \begin{tabular}{|c|c|}
    \hline
      Augmentation Operation & SA (\%)  \\ \hline
     No augmentation & 51.50    \\ \hline
    1  & 74.82    \\ \hline
    2 & 76.86    \\ \hline
    3   & 78.47    \\ \hline
    4  & 76.15    \\ \hline
    12  & 77.96   \\ \hline
    13  & 76.68    \\ \hline
    14  & 71.47    \\ \hline
    23  & 78.98    \\ \hline
    24  & 77.70   \\ \hline
    34  & 77.15    \\ \hline
    123  & 79.94   \\ \hline
    124  & 77.18    \\ \hline
    134  & 76.00    \\ \hline
    234  & 79.43    \\ \hline
    1234  & 78.97   \\ \hline
  \end{tabular}
  \label{impact_of_aug}
\end{table}

\bibliographystyle{ACM-Reference-Format}
\bibliography{refs}

\appendix

\begin{table}[!t]\renewcommand{\arraystretch}{1.2} 
	\centering
	\caption{Impact of the parameters in pre-training a target encoder, training a stolen encoder, and training a downstream classifier on {\name}. The pre-training dataset is CIFAR10 and downstream dataset is GTSRB. }
	\setlength{\tabcolsep}{0.6mm}
	{
	\begin{tabular}{|c|c|c|c|c|c|}
		\hline
	\makecell{Stage} & \makecell{Parameter} & \makecell{Value} & \makecell{TA (\%)} & \makecell{SA (\%)} \\ \hline \hline
	\multirow{6}{*}{\makecell{Pre-training \\ target encoder}}
	&	\multirow{3}{*}{\#epochs} 
	    &  500   & 78.92 & 77.50 \\ \cline{3-5} 
	    && 1,000 & 80.67 & 79.43 \\ \cline{3-5} 
	    && 1,500 & 79.22 & 78.87 \\ \cline{2-5} 
	&	\multirow{3}{*}{\makecell{Learning\\rate}} 
	    & $1\times 10^{-3}$   & 80.67 & 79.43  \\ \cline{3-5}
	    && $5\times 10^{-4}$  & 78.09 & 78.70  \\ \cline{3-5}  
	    && $1\times 10^{-4}$  & 76.83 & 75.03  \\ \hline \hline

	\multirow{9}{*}{\makecell{StolenEncoder}}
	&	\multirow{3}{*}{\#epochs} 
		&	10 & \multirow{9}{*}{80.67} & 51.11 \\ \cline{3-3} \cline{5-5} 
	    && 100 &  & 79.43 \\ \cline{3-3} \cline{5-5} 
	    && 200 &  & 81.74 \\ \cline{2-3} \cline{5-5}  
	&	\multirow{3}{*}{\makecell{Learning\\rate}} 
	    &  $1\times 10^{-3}$ & & 79.43 \\ \cline{3-3} \cline{5-5}  
	    && $5\times 10^{-4}$ & & 76.21 \\ \cline{3-3} \cline{5-5} 
	    && $1\times 10^{-4}$ & & 72.22 \\ \cline{2-3} \cline{5-5}  
	&	\multirow{3}{*}{ Batch size} 
	    &  32 &  & 79.47 \\ \cline{3-3} \cline{5-5}  
	    && 64 &  & 79.43 \\ \cline{3-3} \cline{5-5} 
	    && 128 & & 79.96 \\ \hline \hline
	    
	\multirow{9}{*}{\makecell{Training a \\ downstream \\classifier}}
	&	\multirow{3}{*}{ \#epochs} 
	    &  100 & 65.47 & 62.45 \\ \cline{3-5}  
	    && 300 & 78.01 & 75.28 \\ \cline{3-5}  
	    && 500 & 80.67 & 79.43 \\ \cline{2-5}  
	&	\multirow{3}{*}{\makecell{Learning\\rate}} 
	    &  $1 \times 10^{-3}$ & 73.44 & 71.12 \\ \cline{3-5}  
	    && $5 \times 10^{-4}$ & 81.97 & 80.40 \\ \cline{3-5}  
	    && $1 \times 10^{-4}$ & 80.67 & 79.43  \\ \cline{2-5}  
 
	&	\multirow{3}{*}{\makecell{\#neurons \\in the two \\hidden layers} }
	    &  [128,64]  & 74.58 & 72.56 \\ \cline{3-5}   
	    && [256,128] & 78.04 & 76.48 \\ \cline{3-5}  
	    && [512,256] & 80.67 & 79.43 \\ \hline
	\end{tabular}
	}
	\label{result_of_different_parameters}
%  \vspace{-3mm}
\end{table}

\end{document}